\documentclass{amsart}
\usepackage{graphicx}

\input{tcilatex}
\input tcilatex

\begin{document}
\title[genetic drift, mutation-fitness balance and sampling formulae]{Random evolutionary dynamics driven by fitness and house-of-cards mutations.
Sampling formulae}
\author{Thierry E. Huillet}
\address{Laboratoire de Physique Th\'{e}orique et Mod\'{e}lisation \\
CNRS-UMR 8089 et Universit\'{e} de Cergy-Pontoise, 2 Avenue Adolphe Chauvin,
95302, Cergy-Pontoise, FRANCE\\
E-mail: Thierry.Huillet@u-cergy.fr}
\maketitle

\begin{abstract}
We first revisit the multi-allelic mutation-fitness balance problem,
especially when mutations obey a house of cards condition, where the
discrete-time deterministic evolutionary dynamics of the allelic frequencies derives from
a Shahshahani potential. We then consider multi-allelic Wright-Fisher
stochastic models whose deviation to neutrality is from the Shahshahani
mutation/selection potential. We next focus on the weak selection, weak mutation
cases and, making use of a Gamma calculus, we compute the normalizing
partition functions of the invariant probability densities appearing in
their Wright-Fisher diffusive approximations. Using these results,
Generalized Ewens sampling formulae (ESF) from the equilibrium distributions
are derived. We start treating the ESF in the mixed mutation/selection
potential case and then we restrict ourselves to the ESF in the simpler
house-of-cards mutations only situation. We also address some issues
concerning sampling problems from infinitely-many alleles weak limits.%
\newline

\end{abstract}

\section{Introduction}

Population genetics is concerned with the fate of multi-allelic population
frequencies when various driving `forces' such as selection or mutation are
involved. We will briefly revisit the basics of the deterministic dynamics
arising in discrete-time asexual evolutionary genetics when the origin of
motion is either fitness or mutations or both. We will mostly consider the
multi-allelic diploid case under the Hardy-Weinberg hypothesis. Firstly, we
will consider evolution under general fitness mechanisms (Section $2$), then
we deal with general mutation mechanisms (Section $3$). Some particular
fitness/mutation patterns are discussed in the process. In some cases, the
dynamics driven by fitness only takes the form of a Shahshahani gradient
dynamics, as deriving from a Shahshahani selection potential, \cite{Sh}, 
\cite{Ed}. In Section $4$, we give one way to combine the fitness and the
mutation effects. All these issues are of course part of the standard models
discussed for example in \cite{Ew}, \cite{Bu}, \cite{K}, \cite{K1}, \cite{OB}
and \cite{HM3}. We then focus on a reversible mutation pattern in which
mutation probabilities between any two states only depend on the target
state (the house of cards condition for mutations). When combined with
selection effects, the dynamics driven both by selection and house-of-cards
mutations takes the form of a Shahshahani gradient-like dynamics with drift
deriving from a Shahshahani potential mixing additively the mutation and
selection potentials, \cite{Ho}.

On top of such deterministic dynamics, we add (Section $5$) a random genetic
drift while considering a multi-allelic Wright-Fisher Markov chain whose
deviation to neutrality appears as a drift deriving from the Shahshahani
mutation/selection potential. Some scalings brings this Markov chain into
multi-allelic Wright-Fisher diffusion processes with a unique invariant
probability density describing the joint allelic frequencies at equilibrium, 
\cite{THJ}. We consider the weak selection, weak mutation cases and we
compute the normalizing partition functions of the corresponding invariant
densities.

Generalized Ewens sampling formulae (ESF) from such equilibrium allelic
frequencies are then desirable and these are obtained in Section $6$; they
make use of the explicit expressions of the partition functions just
introduced. Such ancient and long-standing questions go back to \cite{Wri};
see \cite{Wa1} and \cite{Li2}. We start treating the ESF in the mixed
mutation/selection potential case (Section $6.1$) and then we restrict
ourselves to the ESF in the simpler house-of-cards mutations only context
(Section $6.2$). We show that such problems are amenable to the evaluation
of some functionals of skew-symmetric Dirichlet distributed random variables
(rvs) and for this purpose, we make extensive use of a `Gamma-calculus'
precisely designed to evaluate such functionals. We also address some issues
concerning the availability of sampling formulas in some infinitely-many
alleles weak limits.

\section{Evolution driven by fitness: the deterministic point of view}

\subsection{Single locus: the case of a diploid population with $K$ alleles}

We briefly describe the frequency distribution dynamics when fitness (or
selection) only drives the process. We consider diploid populations.

\subsubsection{\textbf{Joint evolutionary dynamics}}

Consider $K$ alleles $A_{k}$, $k=1,...,K$ and let $A_{k}A_{l}$ be the
genotypes attached to a single locus. Let $w_{k,l}\in \left[ 0,1\right] $, $%
k,l=1,...,K$ stand for the absolute fitness of the genotypes $A_{k}A_{l}$.
We shall assume $w_{k,l}=w_{l,k}$ ($w_{k,l}$ being the probability of an $%
A_{k}A_{l}$ surviving to maturity, it is natural to take $w_{k,l}=w_{l,k}$).
Let then $W$ be the symmetric fitness matrix with $k,l-$entry $w_{k,l}$.

Assume the current frequency distribution at generation $r\in \left\{
0,1,2,...\right\} $ of the genotypes $A_{k}A_{l}$ is given by $x_{k,l}.$ Let 
$X$ be the frequencies array with $k,l-$entry $x_{k,l}$, obeying $%
\sum_{k,l}x_{k,l}=1$. The joint evolutionary dynamics in the diploid case is
given by the updating\footnote{%
The symbol $^{+}$ is a common and useful notation to denote the updated
frequency from generation $r$ to $r+1$.}: 
\begin{equation}
x_{k,l}^{+}=x_{k,l}\frac{w_{k,l}}{\omega (X)}\text{ where }\omega
(X)=\sum_{k,l}x_{k,l}w_{k,l},  \label{6}
\end{equation}
where the relative fitness of the genotype $A_{k}A_{l}$ is $w_{k,l}/\omega
\left( X\right) $. The joint dynamics takes on the matrix form: 
\begin{equation*}
X^{+}=\frac{1}{\omega (X)}X\circ W=\frac{1}{\omega (X)}W\circ X,
\end{equation*}
where $\circ $ stands for the (commutative) Hadamard product of matrices.

With $\mathbf{1}$ a column-vector of ones, $\mathbf{1}^{\prime }$ its
transpose and $J=\mathbf{11}^{\prime }$ the $K\times K$ matrix whose entries
are all $1$ (the identity for $\circ $)\footnote{%
In the sequel, a boldface variable, say $\mathbf{x}$, will represent a
column-vector so that its transpose, say $\mathbf{x}^{\prime }$, will be a
row-vector. Similarly, $A^{\prime }$ will stand for the transpose of some
matrix $A$.}, then 
\begin{equation*}
\Delta X:=X^{+}-X=\frac{1}{\omega \left( X\right) }\left( W-\omega \left(
X\right) J\right) \circ X=\frac{1}{\omega \left( X\right) }X\circ \left(
W-\omega \left( X\right) J\right) .
\end{equation*}
Let 
\begin{equation}
\sigma ^{2}(X)=\sum_{k,l=1}^{K}x_{k,l}(w_{k,l}-\omega (X))^{2};\text{ }%
\overline{\sigma }^{2}(X)=\sum_{k,l=1}^{K}x_{k,l}\left( \frac{w_{k,l}}{%
\omega (X)}-1\right) ^{2}=\frac{\sigma ^{2}(X)}{\omega (X)^{2}}  \label{8}
\end{equation}
stand respectively for the genotypic variance in absolute fitness and the
diploid variance in relative fitness. Note that, owing to $%
\sum_{k,l}x_{k,l}=1$ and $\omega (X)=\sum_{k,l}x_{k,l}w_{k,l}$, 
\begin{equation*}
\omega (X)\overline{\sigma }^{2}(X)=\omega (X)\sum_{k,l=1}^{K}x_{k,l}\left( 
\frac{w_{k,l}}{\omega (X)}-1\right) ^{2}=\frac{1}{\omega (X)}%
\sum_{k,l}x_{k,l}w_{k,l}^{2}-\omega (X).
\end{equation*}
The partial increase of the mean fitness (where the change of mean fitness
only comes through changes in the frequencies $x_{k,l}$) is given by 
\begin{equation}
\delta \omega \left( X\right) :=\sum_{k,l}\Delta
x_{k,l}w_{k,l}=\sum_{k,l}x_{k,l}\left( \frac{w_{k,l}^{2}}{\omega (X)}%
-w_{k,l}\right) =\omega (X)\overline{\sigma }^{2}(X)>0,  \label{9}
\end{equation}
with a relative rate of increase: $\delta \omega (X)/\omega (X)=\overline{%
\sigma }^{2}(X)$. The latter result (\ref{9}) constitutes the diploid
version of the $1930$ Fisher fundamental theorem of natural selection for
asexual populations; see (\cite{Ew}, Sections $\left( 2.8,\text{ }2.9\right) 
$, \cite{Price}) and \cite{Oka}) for a deeper insight on its meaning.

\subsubsection{\textbf{Marginal multi-allelic dynamics}}

Assuming a Hardy-Weinberg equilibrium, the frequency distribution at
generation $r,$ say $x_{k,l}$, of the genotypes $A_{k}A_{l}$ is given by: $%
x_{k,l}=x_{k}x_{l}$ where $x_{k}=\sum_{l}x_{k,l}$ is the marginal frequency
of type-$k$ allele $A_{k}$ in a genotypic population. The marginal allelic
frequency vector is $\mathbf{x}=X\mathbf{1}$ ($\mathbf{1}$ is the unit $K$%
-column-vector) and the mean allelic fitness is given by the quadratic form: 
$\omega (\mathbf{x}):=\sum_{k,l}x_{k}x_{l}w_{k,l}=\mathbf{x}^{^{\prime }}W%
\mathbf{x}$. The mean fitness $\omega (\mathbf{x})=\mathbf{x}^{^{\prime }}W%
\mathbf{x}$ is homogeneous of degree $d=2$ in the variables $\mathbf{x}$.
Let 
\begin{equation}
\sigma ^{2}(\mathbf{x})=\sum_{k,l=1}^{K}x_{k}x_{l}\left( w_{k,l}-\omega (%
\mathbf{x})\right) ^{2};\text{ }\overline{\sigma }^{2}\left( \mathbf{x}%
\right) =\sum_{k,l=1}^{K}x_{k}x_{l}\left( \frac{w_{k,l}}{\omega (\mathbf{x})}%
-1\right) ^{2}=\frac{\sigma ^{2}(\mathbf{x})}{\omega (\mathbf{x})^{2}}
\label{10a}
\end{equation}
be respectively the genotypic variance in absolute fitness and the diploid
variance in relative fitness.

If we first define the frequency-dependent marginal fitness of $A_{k}$ by $%
w_{k}(\mathbf{x})=(W\mathbf{x})_{k}:=\sum_{l}w_{k,l}x_{l}$, the marginal
dynamics is given by: 
\begin{equation}
x_{k}^{+}=x_{k}\frac{w_{k}\left( \mathbf{x}\right) }{\omega \left( \mathbf{x}%
\right) }=\frac{1}{\omega \left( \mathbf{x}\right) }x_{k}\left( W\mathbf{x}%
\right) _{k}=:p_{k}\left( \mathbf{x}\right) \text{, }k=1,...,K.  \label{15}
\end{equation}
Letting $D_{\mathbf{x}}:=$diag$\left( x_{k},k=1,...,K\right) ,$ if the
allelic frequency distribution is summarized in the column-vector $\mathbf{x}%
:=x_{k}$, $k=1,...,K$ , (\ref{15}) reads in vector form\footnote{$D_{\mathbf{%
x}}W\mathbf{x}$ is the Schur product $\mathbf{x}\circ W\mathbf{x}$ of vector 
$\mathbf{x}$ and vector $W\mathbf{x.}$ See \cite{K} page 238 for a similar
notational convenience.} 
\begin{equation}
\mathbf{x}^{+}=\frac{1}{\omega (\mathbf{x})}D_{\mathbf{x}}W\mathbf{x}=\frac{1%
}{\omega (\mathbf{x})}D_{W\mathbf{x}}\mathbf{x}=:\mathbf{p}\left( \mathbf{x}%
\right) ,  \label{15a}
\end{equation}
where $\mathbf{p}^{\prime }:=\left( p_{1},...,p_{K}\right) $ maps the $%
\left( K-1\right) -$dimensional simplex 
\begin{equation*}
\mathcal{S}_{K}:=\left\{ \mathbf{x}\geq \mathbf{0}:\text{ }\left| \mathbf{x}%
\right| :=\sum_{k=1}^{K}x_{k}=1\right\} ,
\end{equation*}
into itself.\newline

\textbf{Two particular cases:}\newline

$\left( i\right) $ \emph{The haploid case.} Let $\mathbf{w}:=w_{k}\in \left(
0,1\right] ,$ $k=1,...,K,$ denote the absolute fitnesses of the $K$ alleles
and suppose, without loss of generality, that $0<w_{1}\leq ...\leq w_{K}=1$
(so that allele $A_{K}$ has largest fitness $1$). Let $w\left( \mathbf{x}%
\right) :=\sum_{l}w_{l}x_{l}=\mathbf{w}^{^{\prime }}\mathbf{x}$ denote the
mean fitness of the population. Plugging $W=\mathbf{w1}^{\prime }$ in (\ref
{15a}), the dynamics (\ref{15a}) boils down to 
\begin{equation}
\mathbf{x}^{+}=\mathbf{p}\left( \mathbf{x}\right) =\frac{1}{w\left( \mathbf{x%
}\right) }D_{\mathbf{w}}\mathbf{x=}\frac{1}{w\left( \mathbf{x}\right) }D_{%
\mathbf{x}}\mathbf{w,}  \label{4}
\end{equation}
giving the update of the frequency distribution of alleles in an haploid
population where alleles (and not pairs of alleles) are attached to some
locus. Along (\ref{4}), the absolute mean fitness $w\left( \mathbf{x}\right) 
$ increases. Indeed, with $\Delta w(\mathbf{x}):=w(\mathbf{x}^{+})-w(\mathbf{%
x})$: 
\begin{equation*}
\Delta w(\mathbf{x})=\sum_{k}w_{k}\Delta x_{k}=\sum_{k}w_{k}x_{k}\left( 
\frac{w_{k}}{w(\mathbf{x})}-1\right) =\frac{1}{w(\mathbf{x})}\left(
\sum_{k}w_{k}^{2}x_{k}-w(\mathbf{x})^{2}\right) \geq 0,
\end{equation*}
and it is $>0$ except when $\mathbf{x}\in \mathcal{S}_{K}$ is such that $%
\left\{ k:x_{k}>0\right\} \subseteq \left\{ k:w_{k}=1\right\} $, the set of
alleles with maximal fitness. Such $\mathbf{x}$s are equilibrium states of (%
\ref{4}). In particular, an extremal vector $\mathbf{x}^{\prime }=\mathbf{e}%
_{k}^{\prime }:=\left( 0,...,0,1,0,...,0\right) $ with $k\in \left\{
k:w_{k}=1\right\} $ is a pure (or monomorphic) equilibrium state.

The mean fitness is maximal at equilibrium. The relative rate of increase of 
$w\left( \mathbf{x}\right) $ is: 
\begin{equation}
\frac{\Delta w(\mathbf{x})}{w(\mathbf{x})}=\sum_{k}x_{k}\left( \frac{w_{k}}{%
w(\mathbf{x})}-1\right) ^{2}=\sum_{k}\frac{\left( \Delta x_{k}\right) ^{2}}{%
x_{k}}=\overline{\sigma }^{2}(\mathbf{x}),  \label{5}
\end{equation}
where $\overline{\sigma }^{2}(\mathbf{x})$ is the variance in relative
fitness $\overline{\sigma }^{2}(\mathbf{x})$ given by 
\begin{equation}
\text{ }\overline{\sigma }^{2}(\mathbf{x})=\sum_{k=1}^{K}x_{k}\left( \frac{%
w_{k}}{w(\mathbf{x})}-1\right) ^{2}=\frac{\sigma ^{2}(\mathbf{x})}{w(\mathbf{%
x})^{2}}.  \label{3}
\end{equation}
Thus the population mean fitness is non-decreasing. As a consequence, if
there is an unique allele whose fitness strictly dominates the ones of the
others, starting from any initial state which is not an extremal point of $%
\mathcal{S}_{K}$, the haploid trajectories will ultimately converge to this
fittest state (survival of the fittest allele).\newline

$\left( ii\right) $ \emph{The diploid case with multiplicative fitnesses.}
Suppose that $w_{k,l}=w_{k}w_{l}$, or in matrix form that $W=\mathbf{ww}%
^{\prime }.$ Then selection acts on the gametes rather than on the
genotypes. Observing $\frac{w_{k}\left( \mathbf{x}\right) }{\mathbf{x}%
^{^{\prime }}W\mathbf{x}}=\frac{w_{k}}{\sum_{l}w_{l}x_{l}}$, the dynamics (%
\ref{15a}) boils down to the one (\ref{4}) of haploid populations. However,
the mean fitness in this case is $\omega \left( \mathbf{x}\right) =\left(
\sum_{l}w_{l}x_{l}\right) ^{2}$ and not $w\left( \mathbf{x}\right)
=\sum_{l}w_{l}x_{l}$ as in the haploid case.

\subsubsection{\textbf{Increase of mean fitness for diploid populations}}

Similarly, defining $0\leq R\left( \mathbf{x}\right) :=\sum_{k,l}x_{k}\left(
1-\frac{w_{k}\left( \mathbf{x}\right) }{\omega \left( \mathbf{x}\right) }%
\right) w_{k,l}\left( 1-\frac{w_{l}\left( \mathbf{x}\right) }{\omega \left( 
\mathbf{x}\right) }\right) x_{l}$, in the diploid case we have 
\begin{equation}
\Delta \omega \left( \mathbf{x}\right) =\omega \left( \mathbf{x}^{+}\right)
-\omega \left( \mathbf{x}\right) =R\left( \mathbf{x}\right) +\frac{2}{\omega
\left( \mathbf{x}\right) }\left( \sum_{k}x_{k}w_{k}\left( \mathbf{x}\right)
^{2}-\omega \left( \mathbf{x}\right) ^{2}\right) \geq 0.  \label{18a}
\end{equation}
The mean fitness $\omega \left( \mathbf{x}\right) =\mathbf{x}^{^{\prime }}W%
\mathbf{x}$ for diploid populations, as a Lyapunov function, increases as
time passes by, vanishing only when the process has reached equilibrium.
Equation (\ref{18a}) constitutes the mean fitness increase theorem.

The partial rate of increase of the mean fitness due to frequency shifts
only (see \cite{Ew}) is $\delta \omega \left( \mathbf{x}\right)
:=\sum_{k}\Delta x_{k}w_{k}\left( \mathbf{x}\right) .$ It satisfies 
\begin{equation}
\frac{\delta \omega \left( \mathbf{x}\right) }{\omega \left( \mathbf{x}%
\right) }=\sum_{k=1}^{K}x_{k}\left( \frac{w_{k}\left( \mathbf{x}\right) }{%
\omega \left( \mathbf{x}\right) }-1\right) ^{2}=\sum_{k=1}^{K}\frac{\left(
\Delta x_{k}\right) ^{2}}{x_{k}}=\frac{1}{2}\overline{\sigma }_{A}^{2}\left( 
\mathbf{x}\right) >0,  \label{18aa}
\end{equation}
where $\overline{\sigma }_{A}^{2}\left( \mathbf{x}\right) $ is the allelic
variance in relative fitness 
\begin{equation}
\overline{\sigma }_{A}^{2}\left( \mathbf{x}\right)
:=2\sum_{k=1}^{K}x_{k}\left( \frac{w_{k}\left( \mathbf{x}\right) }{\omega
\left( \mathbf{x}\right) }-1\right) ^{2}.  \label{18ab}
\end{equation}
Equation (\ref{18aa}) constitutes the diploid version of the Fisher
fundamental theorem of natural selection under the Hardy-Weinberg condition
involving random mating. See \cite{Price}, \cite{Ew} and \cite{Cas}.

\subsubsection{\textbf{Shahshahani gradient-like representation of the
allelic dynamics }(\ref{15a})}

There is an alternative vectorial representation of the dynamics (\ref{15a})
emphasizing its gradient-like character. With $\mathbf{x}\in \mathcal{S}_{K}$%
, define the matrix $G(\mathbf{x})=D_{\mathbf{x}}-\mathbf{xx}^{^{\prime }}$.
It is symmetric, positive semi-definite whose quadratic form vanishes only
for the constants and\textbf{\ }$G(\mathbf{e}_{k})=0$\textbf{\ }for all $k$. 
$G\left( \mathbf{x}\right) $ is partially invertible on the space $\Delta
_{K}$ orthogonal to the constants with left-inverse 
\begin{equation*}
G\left( \mathbf{x}\right) ^{-1}=\left( I-\frac{1}{K}J\right) D_{\mathbf{x}%
}^{-1},
\end{equation*}
so with\textbf{\ }$G(\mathbf{x})^{-1}G\left( \mathbf{x}\right) \mathbf{%
\delta }=\mathbf{\delta ,}$ for all $\mathbf{\delta \in }\Delta _{K}$,
obeying\textbf{\ }$\left| \mathbf{\delta }\right| =0$\textbf{.} Note $%
G\left( \mathbf{x}\right) G(\mathbf{x})^{-1}\mathbf{\delta }\neq \mathbf{%
\delta }$ and $\left| G\left( \mathbf{x}\right) \mathbf{\delta }\right| =0$
for all\textbf{\ }$\mathbf{\delta \in }\Delta _{K}$. Looking for a
left-inverse in the weaker sense of the quadratic form, that is satisfying 
\begin{equation*}
\mathbf{\delta }^{^{\prime }}G\left( \mathbf{x}\right) ^{-1}G\left( \mathbf{x%
}\right) \mathbf{\delta }=\mathbf{\delta }^{^{\prime }}I\mathbf{\delta }
\end{equation*}
for all $\mathbf{\delta \in }\Delta _{K}$\textbf{\ }with $\left| \mathbf{%
\delta }\right| =0,$ every $G\left( \mathbf{x}\right) ^{-1}=\left( I-\frac{%
\lambda }{K}J\right) D_{\mathbf{x}}^{-1}$ would do for any real number $%
\lambda $. In particular $\lambda =0,$ leading to\textbf{\ }$G\left( \mathbf{%
x}\right) ^{-1}=D_{\mathbf{x}}^{-1}$. \newline

Introduce the quantity $V_{W}(\mathbf{x})=\frac{1}{2}\log \omega (\mathbf{x}%
) $. Then, (\ref{15}) may be recast as the gradient-like dynamics: 
\begin{equation}
\Delta \mathbf{x}=\frac{1}{\omega (\mathbf{x})}G(\mathbf{x})W\mathbf{x}=G(%
\mathbf{x})\nabla V_{W}(\mathbf{x}),  \label{18c}
\end{equation}
with $\Delta \mathbf{x\in }\Delta _{K},$\ as a result of $\left| \Delta 
\mathbf{x}\right| =\mathbf{1}^{^{\prime }}\Delta \mathbf{x}=0$\ observing%
\textbf{\ }$\mathbf{1}^{^{\prime }}G\left( \mathbf{x}\right) =\mathbf{0}%
^{^{\prime }}$. Note 
\begin{equation*}
\nabla V_{W}\left( \mathbf{x}\right) ^{^{\prime }}\Delta \mathbf{x}=\nabla
V_{W}\left( \mathbf{x}\right) ^{^{\prime }}G\left( \mathbf{x}\right) \nabla
V_{W}\left( \mathbf{x}\right) \geq 0.
\end{equation*}
Based on \cite{Sh}, \cite{Sv}, the dynamics (\ref{18c}) is of gradient-type
with respect to the Shahshahani-Svirezhev metric $G$. Its piecewise
trajectories are perpendicular to the level surfaces of $V_{W}$\ with
respect to the scalar product given by 
\begin{equation*}
\left\langle \mathbf{\delta }_{1}\mathbf{,\delta }_{2}\right\rangle
_{G}=\left( \mathbf{\delta }_{1}^{^{\prime }}G\left( \mathbf{x}\right) ^{-1}%
\mathbf{\delta }_{2}\right) \text{, }\mathbf{\delta }_{1}\mathbf{,\delta }%
_{2}\in \Delta _{K}.
\end{equation*}
We also have 
\begin{equation*}
d_{G}\left( \mathbf{x,x}^{\prime }\right) =\left\langle \Delta \mathbf{%
x,\Delta x}\right\rangle _{G}^{1/2}=\left( \Delta \mathbf{x}^{^{\prime
}}G\left( \mathbf{x}\right) ^{-1}\Delta \mathbf{x}\right) ^{1/2}=\left(
\sum_{k=1}^{K}x_{k}^{-1}\left( \Delta x_{k}\right) ^{2}\right) ^{1/2}.
\end{equation*}
From (\ref{18aa}) and (\ref{18ab}), $d_{G}\left( \mathbf{x,x}^{\prime
}\right) $, which is the length of $\Delta \mathbf{x}$, is also the
square-root of half the allelic variance (the standard deviation) in
relative fitness.

\subsection{Frequency-dependent fitness}

Consider the dynamics (\ref{15a}) in vector form. Let the
frequency-dependent marginal fitness of $A_{k}$ be defined by $w_{k}(\mathbf{%
x})>0$, $k=1,...,K$, not necessarily of the linear form $(W\mathbf{x})_{k}$
for some fitness matrix $W.$ With $\mathbf{w}(\mathbf{x}):=\left( w_{1}(%
\mathbf{x}),...,w_{K}(\mathbf{x})\right) ^{^{\prime }}$ a new column vector
of frequency-dependent marginal fitnesses, we can first think of defining
the dynamics on $\mathcal{S}_{K}$ by 
\begin{equation}
\mathbf{x}^{+}=\frac{1}{\omega (\mathbf{x})}D_{\mathbf{x}}\mathbf{w}(\mathbf{%
x})=:\mathbf{p}\left( \mathbf{x}\right) ,  \label{fd}
\end{equation}
where $\omega (\mathbf{x}):=\mathbf{x}^{^{\prime }}\mathbf{w}(\mathbf{x})$
is the new mean fitness of the allelic population. Unless $\mathbf{w}(%
\mathbf{x})=W\mathbf{x}$ as before, such dynamics cannot be of
Shahshahani-gradient type. This suggests to consider the alternative
gradient-like dynamics to (\ref{fd}), still on the simplex: 
\begin{equation}
\mathbf{x}^{+}=\mathbf{x}+G(\mathbf{x})\nabla V_{W}(\mathbf{x})=:\mathbf{p}%
\left( \mathbf{x}\right) ,  \label{fds}
\end{equation}
where $V_{W}(\mathbf{x})=\frac{1}{2}\log \omega (\mathbf{x})$ and $\omega (%
\mathbf{x}):=\mathbf{x}^{^{\prime }}\mathbf{w}(\mathbf{x})=%
\sum_{k}x_{k}w_{k}(\mathbf{x}).$ A particular case is $\mathbf{w}(\mathbf{x}%
)=W\left( \mathbf{x}\right) \mathbf{x}$ where $W$ is now frequency-dependent
and symmetric for each $\mathbf{x}$ and $\omega (\mathbf{x})=\mathbf{x}%
^{^{\prime }}W\left( \mathbf{x}\right) \mathbf{x}$. Fitness landscapes can
be more general than quadratic forms.

\emph{Examples:}

- Suppose $W\left( \mathbf{x}\right) _{k,l}=\sum_{j=1}^{K}W_{k,l}^{j}x_{j}$
where $W_{k,l}^{j}=W_{l,k}^{j}$ for all $j,k,l\in \left\{ 1,...,K\right\} .$
Then the mean fitness $\omega (\mathbf{x})=\mathbf{x}^{^{\prime }}W\left( 
\mathbf{x}\right) \mathbf{x}$ is homogeneous of degree $d=3$ in the
variables $\mathbf{x}$.

- With $\sigma ,q>0$, let $\mathbf{\sigma }_{q}\left( \mathbf{x}\right)
=\sigma \left( x_{1}^{q-2},...,x_{K}^{q-2}\right) ^{\prime }$ and suppose $%
W\left( \mathbf{x}\right) =D_{\mathbf{\sigma }_{q}\left( \mathbf{x}\right) }$
is diagonal, so that $\omega (\mathbf{x})=\sigma \sum_{k=1}^{K}x_{k}^{q}$.
Such selection models were considered in \cite{Wa1}, \cite{Eli}, \cite{Han}, 
\cite{H1} and \cite{CJ}, in relation to heterozygozity. $\Diamond $

\section{Evolution driven by mutation}

We now briefly describe the frequency distribution dynamics when mutation is
the only driving source of motion.\newline
Assume alleles mutate according to the scheme: $A_{k}\rightarrow A_{l}$ with
probability $\mu _{k,l}\in \left[ 0,1\right] $ satisfying $\mu _{k,k}=0$ and 
$0<\sum_{l\neq k}\mu _{k,l}\leq 1$ for all $k.$ Let $M:=\left[ \mu
_{k,l}\right] $ be the mutation pattern matrix; we shall assume that the
non-negative matrix $M$ is irreducible.

\subsection{Frequency dynamics under mutation only}

Considering first an updating mechanism of the frequencies where only
mutations operate, we get 
\begin{equation}
x_{k}^{+}=x_{k}+\sum_{l\neq k}\mu _{l,k}x_{l}-x_{k}\sum_{l\neq k}\mu _{k,l}%
\text{, }k=1,...,K,  \label{34}
\end{equation}
whose meaning is the one of a ``master equation''. In vector form, with $%
M^{^{\prime }}$ the transpose of $M$ 
\begin{equation}
\mathbf{x}^{+}=\mathbf{x+}M^{^{\prime }}\mathbf{x-}D_{M\mathbf{1}}\mathbf{x}%
=:\mathbf{Mx}=:\mathbf{p}_{M}\left( \mathbf{x}\right) \mathbf{,}  \label{35}
\end{equation}
and the update of the frequencies with mutations is given by the linear
transformation 
\begin{equation*}
\mathbf{M}:=I\mathbf{-}D_{M\mathbf{1}}+M^{^{\prime }}.
\end{equation*}
The vector $\mathbf{m}:=M\mathbf{1}$\ is called the mutation load. We have $%
\mathbf{M}\geq \mathbf{0}$ and $\mathbf{M}=M^{^{\prime }}$ if and only if $M$
is stochastic: $\mathbf{m}=M\mathbf{1}=\mathbf{1,}$meaning 
\begin{equation*}
\sum_{l\neq k}\mu _{k,l}=1\text{ for all }k.
\end{equation*}
Also $\mathbf{1}^{^{\prime }}\mathbf{M}=\mathbf{1}^{^{\prime }}$ and then $%
\mathbf{M}$ maps $\mathcal{S}_{K}$ into $\mathcal{S}_{K}$ because if $%
\mathbf{1}^{^{\prime }}\mathbf{x}=1$, then $\mathbf{1}^{^{\prime }}\mathbf{x}%
^{+}=\mathbf{1}^{^{\prime }}\mathbf{Mx}=\left( \mathbf{M}^{^{\prime }}%
\mathbf{1}\right) ^{^{\prime }}\mathbf{x}=\mathbf{1}^{^{\prime }}\mathbf{x}%
=1 $. The matrix $\mathbf{M}^{^{\prime }}$ is stochastic and irreducible and
so, by Perron-Frobenius theorem, it has a unique strictly positive
probability left-eigenvector associated to the real dominant eigenvalue $1.$
Let $\mathbf{x}_{eq}^{\prime }$ be this row-vector, so obeying $\mathbf{x}%
_{eq}^{\prime }=\mathbf{x}_{eq}^{\prime }\mathbf{M}^{^{\prime }}$, or $%
\mathbf{x}_{eq}=\mathbf{Mx}_{eq}$. Under the irreducibility assumption on $M$%
, the frequencies dynamics involving only mutations has a unique polymorphic
equilibrium fixed point $\mathbf{x}_{eq}>\mathbf{0}$. When $\mathbf{M}$ is
primitive then $\lim_{r\to \infty }\mathbf{M}^{r}=\mathbf{x}_{eq}\mathbf{1}%
^{^{\prime }}.$ This shows that, at generation $r$, 
\begin{equation*}
\mathbf{x}(r)=\mathbf{M}^{r}\mathbf{x}(0)\underset{r\rightarrow \infty }{%
\rightarrow }\mathbf{x}_{eq}\mathbf{1}^{^{\prime }}\mathbf{x}(0)=\mathbf{x}%
_{eq},
\end{equation*}
regardless of the initial condition $\mathbf{x}(0)$ belonging to $\mathcal{S}%
_{K}$. The equilibrium vector $\mathbf{x}_{eq}$\ is asymptotically stable.
These considerations are the same as limit (ergodic) theorems for Markov
chains.

Note finally that from (\ref{35}): 
\begin{equation}
\Delta \mathbf{x}=\left( \mathbf{M}-I\right) \mathbf{x}=:\nabla \mathcal{V}%
_{M}(\mathbf{x}),  \label{36a}
\end{equation}
where $\mathcal{V}_{M}(\mathbf{x})=\frac{1}{2}\mathbf{x}^{^{\prime }}\left( 
\mathbf{M}-I\right) \mathbf{x}$ is the quadratic mutation potential. The
probability right-eigenvector $\mathbf{x}_{eq}$ of $\mathbf{M}$ uniquely
solves $\nabla \mathcal{V}_{M}\left( \mathbf{x}\right) =0$ with $\mathcal{V}%
_{M}(\mathbf{x}_{eq})=0,$ maximal$.$

\subsection{Special mutation patterns}

There are many special mutation scenarii which deserve interest.\newline

$(i)$ Reversible mutations: Let $\mathbf{x}_{eq}$ solve $\mathbf{x}%
_{eq}^{\prime }=\mathbf{x}_{eq}^{\prime }\mathbf{M}^{^{\prime }}.$ Define 
\begin{equation*}
\overleftarrow{\mathbf{M}^{^{\prime }}}=D_{\mathbf{x}_{eq}}^{-1}\mathbf{M}D_{%
\mathbf{x}_{eq}}^{{}}.
\end{equation*}
We have $\overleftarrow{\mathbf{M}^{^{\prime }}}\mathbf{1}=D_{\mathbf{x}%
_{eq}}^{-1}\mathbf{Mx}_{eq}=\mathbf{1}$, so $\overleftarrow{\mathbf{M}%
^{^{\prime }}}$ is the stochastic matrix of the time-reversed mutation
process at equilibrium with invariant measure $\mathbf{x}_{eq}^{\prime }>%
\mathbf{0}^{\prime }$. If $\overleftarrow{\mathbf{M}^{\prime }}=\mathbf{M}%
^{^{\prime }}$, then the mutation pattern is said to be time-reversible
(detailed balance holds). In this case 
\begin{equation*}
\mu _{l,k}=\mu _{k,l}\frac{x_{eq,k}}{x_{eq,l}},
\end{equation*}
with $\mu _{k,l}>0\Rightarrow \mu _{l,k}>0$. When reversible, the matrix $%
\mathbf{M}^{^{\prime }}$\ is diagonally similar to the matrix $\mathbf{M}%
_{S}^{^{\prime }}:=D_{\mathbf{x}_{eq}}^{1/2}\mathbf{M}^{^{\prime }}D_{%
\mathbf{x}_{eq}}^{-1/2}$\ which is symmetric with real eigenvalues, so $%
\mathbf{M}^{\prime }$ and $\mathbf{M}$ have real eigenvalues.\newline

$(ii)$ If $M=M^{^{\prime }}$, then $\mathbf{M}=\mathbf{M}^{^{\prime }}$ and $%
\mathbf{M}$ is symmetric itself and thus doubly stochastic. In that case, $%
\mathbf{x}_{eq}=\frac{1}{K}\cdot \left( 1,...,1\right) ^{^{\prime }}=:%
\mathbf{x}_{b}$ (the barycenter of $\mathcal{S}_{K}$) and such mutation
patterns are reversible. Let us give some \emph{Examples}:

- A model with symmetric mutations is obtained for instance while assuming
multiplicative mutations: $\mu _{k,l}=\mu _{k}\mu _{l}.$ In this case, with $%
\mathbf{\mu }$ the column vector of the $\mu _{k}$s, $k=1,...,K,$%
\begin{equation*}
\mathbf{M}=I+\mathbf{\mu \mu }^{^{\prime }}-\left| \mathbf{\mu }\right| D_{%
\mathbf{\mu }}.
\end{equation*}
This mutation pattern is reversible with $\mathbf{x}_{eq}=\mathbf{x}_{b}$.

- Alternatively, assuming $\mu _{k,l}\equiv \mu \in \left( 0,\frac{1}{K-1}%
\right] $ for all $k\neq l$ leads to $\mathbf{M}=\mu J+\left( 1-K\mu \right)
I$ which is also symmetric.

- Alternatively, while considering additive mutations: $\mu _{k,l}=\left(
\mu _{k}+\mu _{l}\right) /2,$\ with $\mathbf{\mu }^{^{\prime }}=\left( \mu
_{1},...,\mu _{K}\right) $, 
\begin{equation*}
\mathbf{M}=\left( 1-\frac{1}{2}\left| \mathbf{\mu }\right| \right) I-\frac{K%
}{2}D_{\mathbf{\mu }}+\frac{1}{2}\left( \mathbf{\mu 1}^{\prime }+\mathbf{%
1\mu }^{^{\prime }}\right)
\end{equation*}
and $\mathbf{x}_{eq}=\mathbf{x}_{b}$. This mutation pattern is reversible.%
\newline

$\left( iii\right) $ It is not necessary that $M=M^{^{\prime }}$ in order to
have $\mathbf{M}$ doubly stochastic. It suffices to impose $M\mathbf{1}%
=M^{^{\prime }}\mathbf{1}$. In that case although $\mathbf{M}\neq \mathbf{M}%
^{^{\prime }},$ the overall input-output mutation probabilities attached to
any state coincide and the equilibrium state again matches with the
barycenter $\mathbf{x}_{b}$ of $\mathcal{S}_{K}$. But since $M\neq
M^{^{\prime }}$, such mutation patterns are not reversible.\newline

$(iv)$ Assume the mutation probabilities only depend on the initial state,
that is: $\mu _{k,l}=\mu _{k}$ for all $l\neq k$. Then 
\begin{equation*}
\mathbf{M}=I-KD_{\mathbf{\mu }}+\mathbf{1\mu }^{^{\prime }}.
\end{equation*}
This mutation model is not reversible because 
\begin{equation*}
\overleftarrow{\mathbf{M}^{^{\prime }}}=D_{\mathbf{\mu }}^{-1}\mathbf{M}D_{%
\mathbf{\mu }}^{{}}=I-KD_{\mathbf{\mu }}+D_{\mathbf{\mu }}^{-1}\mathbf{1\mu }%
^{^{\prime }}D_{\mathbf{\mu }}^{{}}\neq \mathbf{M}^{^{\prime }}=I-KD_{%
\mathbf{\mu }}+\mathbf{\mu 1}^{^{\prime }}.
\end{equation*}
If $\mu _{k}>0$\ for all $k$, the equilibrium state is 
\begin{equation*}
x_{eq,k}=\frac{1/\mu _{k}}{\sum_{l}1/\mu _{l}}\text{ or }\mathbf{x}_{eq}=%
\frac{1}{\text{trace}\left( D_{\mathbf{\mu }}^{-1}\right) }D_{\mathbf{\mu }%
}^{-1}\mathbf{1}.
\end{equation*}
\newline

$\left( v\right) $ One-way (irreversible) mutations: assume $\mu
_{k,l}>0\Rightarrow \mu _{l,k}=0.$\ This model is clearly not reversible
and, when the associated mutation matrix $M$\ is irreducible, it has a
non-trivial $\mathbf{x}_{eq}>\mathbf{0}.$ This model includes the cyclic
mutation pattern for which $\mu _{k,l}=\mu _{k}\delta _{l,k+1}$, $%
k=1,...,K-1 $ and $\mu _{K,l}=\mu _{K}\delta _{l,1},$ with 
\begin{equation*}
\mathbf{x}_{eq}=\frac{1}{\text{trace}\left( D_{\mathbf{\mu }}^{-1}\right) }%
D_{\mathbf{\mu }}^{-1}\mathbf{1},
\end{equation*}
as in the previous example $(iv)$.\newline

$(vi)$ (Kingman house-of-cards mutations, \cite{K1}).

Assume the mutation probabilities now only depend on the terminal state,
that is: $\mu _{k,l}=\mu _{l}$ for all $k\neq l$, still with $\mu _{k,k}=0.$
Throughout, we assume $\mu _{k}>0$. Let $\mathbf{\mu }^{^{\prime }}=\left(
\mu _{1},...,\mu _{K}\right) $. Then, $M=\mathbf{1\mu }^{^{\prime }}-D_{%
\mathbf{\mu }}$, $M\mathbf{1}=\left| \mathbf{\mu }\right| \mathbf{\cdot 1}-%
\mathbf{\mu }$ where $\min \mu _{k}<\left| \mathbf{\mu }\right| :=\mathbf{%
\mu }^{^{\prime }}\mathbf{1}<1+\max \mu _{k}$, $\mathbf{M}=\mathbf{\mu 1}%
^{^{\prime }}+\left( 1-\left| \mathbf{\mu }\right| \right) I$ and 
\begin{equation}
\mathbf{x}^{+}=\mathbf{Mx}=\mathbf{x+}M^{^{\prime }}\mathbf{x-}D_{\mathbf{x}%
}M\mathbf{1}=\mathbf{\mu }+\left( 1-\left| \mathbf{\mu }\right| \right) 
\mathbf{x.}  \label{37}
\end{equation}
The equilibrium state is $\mathbf{x}_{eq}=\mathbf{\mu }/\left| \mathbf{\mu }%
\right| $ and it is stable. Note that $\left| \mathbf{\mu }\right| \leq 1+%
\frac{1}{K-1}$. This mutation model is reversible because 
\begin{equation*}
\overleftarrow{\mathbf{M}^{^{\prime }}}=D_{\mathbf{\mu }}^{-1}\mathbf{M}D_{%
\mathbf{\mu }}^{{}}=D_{\mathbf{\mu }}^{-1}\mathbf{\mu 1}^{^{\prime }}D_{%
\mathbf{\mu }}^{{}}+\left( 1-\left| \mathbf{\mu }\right| \right) I=\mathbf{%
1\mu }^{^{\prime }}+\left( 1-\left| \mathbf{\mu }\right| \right) I=\mathbf{M}%
^{^{\prime }}.
\end{equation*}
In this model the coordinates are decoupled: $x_{k}^{+}=\mu _{k}+\left(
1-\left| \mathbf{\mu }\right| \right) x_{k}$, depends only on $x_{k}$. We
shall come back at length to this mutation pattern in the sequel.\newline

- Note that if $\mathbf{m}=M\mathbf{1}=\mathbf{1}$, namely if $\left( \left| 
\mathbf{\mu }\right| -1\right) \mathbf{\cdot 1}=\mathbf{\mu }$, then $%
\mathbf{x}_{eq}=\mathbf{x}_{b}$, $K\left( \left| \mathbf{\mu }\right|
-1\right) =\left| \mathbf{\mu }\right| $ (else $\left| \mathbf{\mu }\right|
-1=1/\left( K-1\right) $) and 
\begin{equation*}
\mathbf{x}^{+}=\mathbf{\mu }+\left( 1-\left| \mathbf{\mu }\right| \right) 
\mathbf{x}=\left( \left| \mathbf{\mu }\right| -1\right) \left( \mathbf{1-x}%
\right) =\frac{1}{K-1}\left( \mathbf{1-x}\right) .
\end{equation*}

- In the latter case, $\left| \mathbf{\mu }\right| >1$\ in particular. If $%
\left| \mathbf{\mu }\right| >1$ and $\mathbf{m}\neq \mathbf{1}$, in view of $%
\Delta \mathbf{x}=\left| \mathbf{\mu }\right| \left( \mathbf{x}_{eq}-\mathbf{%
x}\right) $, $\mathbf{x}$ goes fast to $\mathbf{x}_{eq}$.

- Another very special case is when $\left| \mathbf{\mu }\right| =1$. Here $%
\mathbf{M}=\mathbf{\mu 1}^{^{\prime }}$\ and $\mathbf{x}^{+}=\mathbf{Mx=\mu }
$. Starting from any initial condition, the dynamics moves in one-step to $%
\mathbf{x}_{eq}=\mathbf{\mu }$\ (inside the simplex $\mathcal{S}_{K}$) and
stays there for ever. $\Diamond $

\section{Evolution driven by combined fitness and mutation forces}

Let us now consider the dynamics driven both by fitness and mutation.

\subsection{The combined fitness/mutation frequency dynamics; \protect\cite
{Ho}, \protect\cite{OB}, \protect\cite{H}, \protect\cite{HM3}}

Combining the fitness and mutation effects consists in applying first the
fitness mapping and then let mutation act on the result. Proceeding in this
way, we get the `fitness-first' dynamics \cite{Ho}: 
\begin{equation}
\mathbf{x}^{+}=\frac{1}{\mathbf{x}^{^{\prime }}W\mathbf{x}}\mathbf{M}D_{W%
\mathbf{x}}\mathbf{x}=\frac{1}{\mathbf{x}^{^{\prime }}W\mathbf{x}}\mathbf{M}%
D_{\mathbf{x}}W\mathbf{x,}  \label{38}
\end{equation}
defining a new nonlinear transformation. Alternatively, $\mathbf{x}^{+}=%
\mathbf{p}\left( \mathbf{x}\right) $ where $\mathbf{p}\left( \mathbf{x}%
\right) =\frac{1}{\mathbf{x}^{^{\prime }}W\mathbf{x}}\mathbf{M}D_{\mathbf{x}%
}W\mathbf{x}$ is the new mapping from $\mathcal{S}_{K}$ to $\mathcal{S}_{K}$
to consider. Component-wise, this is also as required 
\begin{equation}
x_{k}^{+}=\frac{1}{\omega (\mathbf{x})}\left( x_{k}w_{k}(\mathbf{x})\mathbf{+%
}\sum_{l\neq k}\mu _{l,k}w_{l}(\mathbf{x})x_{l}-x_{k}w_{k}(\mathbf{x}%
)\sum_{l\neq k}\mu _{k,l}\right) \text{, }k=1,...,K.  \label{39}
\end{equation}
We have: $\mathbf{p}(\mathbf{e}_{k})=\left( \mu _{k,1},...,\mu
_{k,k-1},1-\sum_{l\neq k}\mu _{k,l},\mu _{k,k+1},...,\mu _{k,K}\right)
^{^{\prime }}\in \mathcal{S}_{K}$ and so the extremal states $\mathbf{e}_{k}$
are not invariant under $\mathbf{p}$ and from the fixed-point theorem, there
exists some equilibrium state in $\mathcal{S}_{K}$. Using the representation
(\ref{18c}), (\ref{35}) and (\ref{36a}): 
\begin{equation}
\Delta \mathbf{x}=\left( \mathbf{M}-I\right) \mathbf{x}+\mathbf{M}G(\mathbf{x%
})\nabla V_{W}(\mathbf{x})=\nabla \mathcal{V}_{M}(\mathbf{x})+\mathbf{M}G(%
\mathbf{x})\nabla V_{W}(\mathbf{x}).  \label{40}
\end{equation}
This is not a Shahshahani gradient-like dynamics in general. \cite{Ho} also
considers a continuous-time version of (\ref{40}).

When $\mathbf{M}=I$ (no mutation) (\ref{40}) boils down into (\ref{18c}) and
when $W=J$ (no selection), (\ref{40}) boils down into (\ref{36a}).

When both $\mathbf{M}=I$ (no mutation) and $W=J$ (no selection), $\Delta 
\mathbf{x}=0$ with corresponding neutral $\mathbf{p}\left( \mathbf{x}\right)
=\mathbf{x}.$

\subsection{Fitness/mutation frequency dynamics in the house of cards
condition for mutations}

This is a remarkable case where the allelic dynamics driven both by fitness
and mutation forces has a Shahshahani gradient-like structure, \cite{Ho}.
Indeed, from (\ref{37}), $\mathbf{M}=\mathbf{\mu 1}^{^{\prime }}+\left(
1-\left| \mathbf{\mu }\right| \right) I$\ and (\ref{38}) boils down to 
\begin{equation*}
\mathbf{x}^{+}=\mathbf{\mu +}\left( 1-\left| \mathbf{\mu }\right| \right) 
\frac{1}{\mathbf{x}^{^{\prime }}W\mathbf{x}}D_{W\mathbf{x}}\mathbf{x}.
\end{equation*}
We now have 
\begin{equation*}
\left( \mathbf{M}-I\right) \mathbf{x=\mu }-\left| \mathbf{\mu }\right| 
\mathbf{x}=G(\mathbf{x})\nabla V_{M}(\mathbf{x}),
\end{equation*}
where $V_{M}(\mathbf{x})=\sum_{k}\mu _{k}\log x_{k}-\left| \mathbf{\mu }%
\right| \sum_{k}x_{k}$. Indeed, $\nabla V_{M}\left( \mathbf{x}\right) $\ $=$%
\ $D_{\mathbf{x}}^{-1}\mathbf{\mu }-\left| \mathbf{\mu }\right| \mathbf{1}$\
and 
\begin{equation*}
\left( D_{\mathbf{x}}-\mathbf{xx}^{^{\prime }}\right) \nabla V_{M}\left( 
\mathbf{x}\right) =\mathbf{\mu }-\left| \mathbf{\mu }\right| \mathbf{x}\text{%
.}
\end{equation*}
Furthermore, 
\begin{equation*}
\mathbf{M}G(\mathbf{x})=\left( \mathbf{\mu 1}^{^{\prime }}+\left( 1-\left| 
\mathbf{\mu }\right| \right) I\right) \left( D_{\mathbf{x}}-\mathbf{xx}%
^{^{\prime }}\right) =\left( 1-\left| \mathbf{\mu }\right| \right) \left( D_{%
\mathbf{x}}-\mathbf{xx}^{^{\prime }}\right) .
\end{equation*}
Thus (\ref{40}) can alternatively be written as 
\begin{equation}
\Delta \mathbf{x}=G(\mathbf{x})\left( \nabla V_{M}(\mathbf{x})+\left(
1-\left| \mathbf{\mu }\right| \right) \nabla V_{W}(\mathbf{x})\right) =:G(%
\mathbf{x})\nabla V(\mathbf{x})  \label{40pot}
\end{equation}
which is of Shahshahani gradient-type with combined mutation/selection
potential 
\begin{equation}
\begin{array}{l}
V(\mathbf{x})=V_{M}(\mathbf{x})+\left( 1-\left| \mathbf{\mu }\right| \right)
V_{W}(\mathbf{x}) \\ 
=\sum_{k}\mu _{k}\log x_{k}-\left| \mathbf{\mu }\right| +\frac{1}{2}\left(
1-\left| \mathbf{\mu }\right| \right) \log \mathbf{x}^{\prime }W\mathbf{x}
\\ 
=:\log \mathcal{W}(\mathbf{x})\text{, where }\mathcal{W}\left( \mathbf{x}%
\right) =e^{-\left| \mathbf{\mu }\right| }\prod_{k=1}^{K}x_{k}^{\mu
_{k}}\left( \mathbf{x}^{\prime }W\mathbf{x}\right) ^{\left( 1-\left| \mathbf{%
\mu }\right| \right) /2}.
\end{array}
\label{smpot}
\end{equation}
Note that $\mathcal{W}\left( \mathbf{x}\right) $\ is homogeneous of degree $%
1 $.

\subsection{House of cards and polymorphism}

If some polymorphic state $\mathbf{x}_{eq}$\ exists in the interior of the
simplex, then $\mathbf{x}_{eq}=\mathbf{z}/\left| \mathbf{z}\right| $\ where $%
\mathbf{z>0}$\ solves 
\begin{eqnarray*}
\nabla V(\mathbf{z}) &=&D_{\mathbf{z}}^{-1}\mathbf{\mu }+\left( 1-\left| 
\mathbf{\mu }\right| \right) \frac{W\mathbf{z}}{\mathbf{z}^{\prime }W\mathbf{%
z}}=\lambda \mathbf{1,}\text{ else} \\
\lambda \mathbf{z} &=&\mathbf{\mu }+\left( 1-\left| \mathbf{\mu }\right|
\right) \frac{D_{\mathbf{z}}W\mathbf{z}}{\mathbf{z}^{\prime }W\mathbf{z}}%
\mathbf{,}
\end{eqnarray*}
for some arbitrary Lagrangian parameter $\lambda $ ($\mathbf{z}$\ is an
extremum of $V$\ under the constraint $\left| \mathbf{z}\right| $\ fixed).

- If $\mathbf{\mu =0}$\ (no mutation), the searched $\mathbf{z=z}_{S}>%
\mathbf{0}$\ is the one solving, as required, $W\mathbf{z}_{S}\mathbf{%
=\lambda }\left( \mathbf{z}_{S}^{\prime }W\mathbf{z}_{S}\right) \mathbf{1}$\
up to a multiplicative constant, else $W\mathbf{z}_{S}\mathbf{=1}$\ if $%
\lambda =1/\left| \mathbf{z}_{S}\right| \mathbf{.}$

- If $W=J$\ (no selection), $\frac{D_{\mathbf{z}}W\mathbf{z}}{\mathbf{z}%
^{\prime }W\mathbf{z}}=\frac{\mathbf{z}}{\left| \mathbf{z}\right| }$\ and
the searched $\mathbf{z=z}_{M}>\mathbf{0}$\ is the one solving, as required, 
$\lambda \mathbf{z}_{M}\mathbf{=\mathbf{\mu }+\left( 1-\left| \mathbf{\mu }%
\right| \right) z}_{M}/\left| \mathbf{z}_{M}\right| $\ up to a
multiplicative constant so that $\mathbf{z}_{M}=\mathbf{\mu }$ if $\lambda
=1+\left( 1-\left| \mathbf{\mu }\right| \right) /\left| \mathbf{z}%
_{M}\right| $.

So $\mathbf{x}_{eq}=\mathbf{z}/\left| \mathbf{z}\right| $\ where $\mathbf{z>0%
}$\ is some fixed point of the map 
\begin{equation*}
\mathbf{x\rightarrow }T_{\mathbf{\mu }}\left( \mathbf{x}\right) =\lambda
^{-1}\left( \mathbf{\mu }+\left( 1-\left| \mathbf{\mu }\right| \right) \frac{%
D_{\mathbf{x}}W\mathbf{x}}{\mathbf{x}^{\prime }W\mathbf{x}}\right) .
\end{equation*}
If there is a stable fixed point $\mathbf{z}_{S}>0$ of $T_{\mathbf{0}}$,
such that $W\mathbf{z}_{S}=\mathbf{1}$\ (the purely selection dynamics has a
stable polymorphic equilibrium state $\mathbf{z}_{S}/\left| \mathbf{z}%
_{S}\right| $\ in the simplex), then clearly, as long as $\left| \mathbf{\mu 
}\right| \leq 1$, $\mathbf{z>0}$, as a fixed point of $T_{\mathbf{\mu }}$,
exists as well and is unique, as a maximum of the concave potential $V(%
\mathbf{x})$. If $\mathbf{z}_{S}/\left| \mathbf{z}_{S}\right| $\ is stable
indeed, then so is $\mathbf{x}_{eq}=\mathbf{z}/\left| \mathbf{z}\right| $ if 
$\left| \mathbf{\mu }\right| \leq 1$, because the new mean fitness function $%
w_{\mathbf{\mu }}\left( \mathbf{x}\right) :=\mathcal{W}\left( \mathbf{x}%
\right) $\ inherits the concavity of $\mathbf{x}^{\prime }W\mathbf{x=:}w_{%
\mathbf{0}}\left( \mathbf{x}\right) =w\left( \mathbf{x}\right) $ if $\left| 
\mathbf{\mu }\right| <1$ and is concave as well if $\left| \mathbf{\mu }%
\right| =1$. If $\left| \mathbf{\mu }\right| =1$, then $\mathbf{x}_{eq}=%
\mathbf{\mu }/\left| \mathbf{\mu }\right| $. Note however that $\left| 
\mathbf{\mu }\right| >1$ entails that a stable $\mathbf{z}_{S}/\left| 
\mathbf{z}_{S}\right| $\ can be switched to an unstable $\mathbf{z}/\left| 
\mathbf{z}\right| $: strong mutations ($\left| \mathbf{\mu }\right| >1$) can
destroy a stable polymorphic state of the selection equation.

\section{Evolutionary dynamics: the stochastic point of view}

We now consider a stochastic version of evolutionary dynamics biased by
selection/mutation effects, thereby adding a ``random genetic drift'' to the
deterministic dynamics.

\subsection{Random genetic drift and the Wright-Fisher model}

We will consider the multi-allelic Wright-Fisher model with bias, (see \cite
{Ew}, section $5.10$).

\subsubsection{\textbf{The discrete-time model and first properties}}

Consider a multi-allelic population with constant size $N.$ In the haploid
(diploid) case, $N$ is (twice) the number $N_{e}$ of effective loci. Let $%
\mathbf{n}:=\left( n_{k};\text{ }k=1,...,K\right) $ and $\mathbf{n}%
^{+}:=\left( n_{k}^{+};k=1,...,K\right) $ be two vectors of integers
quantifying the size of the allelic populations at two consecutive
generations $r\varepsilon $ and $r\varepsilon +\varepsilon ,$ where $%
\varepsilon >0$ is some small parameter fixing the time elapsed between two
consecutive generations. With $\left| \mathbf{n}\right| =\sum_{k}n_{k}=N$,
therefore $\left| \frac{\mathbf{n}}{N}\right| =\left| \frac{\mathbf{n}%
^{\prime }}{N}\right| =1$ and both $\frac{\mathbf{n}}{N}$ and $\frac{\mathbf{%
n}^{\prime }}{N}$ belong to $\mathcal{S}_{K}$ $.$ Suppose the stochastic
evolutionary dynamics is now given by a Markov chain whose one-step
probability transition matrix $P$ from state $\mathbf{N}=\mathbf{n}$ to
state $\mathbf{N}^{+}=\mathbf{n}^{+}$ is given by the multinomial
Wright-Fisher model 
\begin{equation}
\mathbf{P}\left( \mathbf{N}_{\left( r+1\right) \varepsilon }=\mathbf{n}%
^{+}\mid \mathbf{N}_{r\varepsilon }=\mathbf{n}\right) =:P\left( \mathbf{n},%
\mathbf{n}^{+}\right) =\binom{N}{n_{1}^{+}\cdots n_{K}^{+}}%
\prod_{k=1}^{K}p_{k}\left( \frac{\mathbf{n}}{N}\right) ^{n_{k}^{+}}.
\label{21}
\end{equation}
Here the $p_{k}$s are the coordinates of some mapping $\mathbf{p}:\mathcal{S}%
_{K}\rightarrow \mathcal{S}_{K},$ translating some bias from neutrality
(where $\mathbf{p}$ is simply the identity). The state-space dimension of
this Markov chain is $\binom{N+K-1}{K-1}$ (the number of compositions of
integer $N$ into $K$ non-negative parts). In other words, with $\mathbf{x:=}%
\frac{\mathbf{n}}{N}\in \mathcal{S}_{K}$, 
\begin{equation*}
\left( \mathbf{N}_{\left( r+1\right) \varepsilon }=\mathbf{n}^{\prime }\mid 
\mathbf{N}_{r\varepsilon }=\mathbf{n}\right) \mathbf{\sim }\text{multinomial}%
\left( N,\mathbf{p}\left( \mathbf{x}\right) \right) .
\end{equation*}

\subsubsection{\textbf{Scalings}}

If we assume that $\mathbf{p}\left( \mathbf{x}\right) -\mathbf{x}%
=\varepsilon \mathbf{f}\left( \mathbf{x}\right) $ for some drift function $%
\mathbf{f}\left( \mathbf{x}\right) $ (meaning $\Delta \mathbf{x}=\varepsilon 
\mathbf{f}\left( \mathbf{x}\right) $ is the frequency shift per generation
of duration $\Delta t=\varepsilon $), from the mean and covariance structure
of multinomial distributions, with $t=r\varepsilon $, $\Delta \mathbf{N}%
_{t}=\left( \mathbf{N}_{t+\varepsilon }-\mathbf{N}_{t}\right) $, we get 
\begin{eqnarray*}
\mathbf{E}_{\mathbf{x}}\left( \Delta \mathbf{N}_{t}\right) &=&\varepsilon N%
\mathbf{f}\left( \mathbf{x}\right) \\
\sigma _{\mathbf{x}}^{2}\left( \Delta \mathbf{N}_{t}\right) &=&\varepsilon
N\left( D_{\mathbf{p}\left( \mathbf{x}\right) }\mathbf{-p}\left( \mathbf{x}%
\right) \mathbf{p}\left( \mathbf{x}\right) ^{^{\prime }}\right) .
\end{eqnarray*}
With $\mathbf{x}_{t}:=\mathbf{N}_{t}/N$ and $\Delta \mathbf{x}_{t}:=\mathbf{x%
}_{t+\varepsilon }-\mathbf{x}_{t}=\left( \mathbf{N}_{t+\varepsilon }-\mathbf{%
N}_{t}\right) /N,$ to the first-order in $\varepsilon $, 
\begin{eqnarray*}
\mathbf{E}_{\mathbf{x}}\left( \Delta \mathbf{x}_{t}\right) &=&\varepsilon 
\mathbf{f}\left( \mathbf{x}\right) \\
\sigma _{\mathbf{x}}^{2}\left( \Delta \mathbf{x}_{t}\right) &=&\varepsilon 
\frac{N}{N^{2}}\left( D_{\mathbf{p}\left( \mathbf{x}\right) }\mathbf{-p}%
\left( \mathbf{x}\right) \mathbf{p}\left( \mathbf{x}\right) ^{^{\prime
}}\right) =\frac{\varepsilon }{N}\left( D_{\mathbf{x}}\mathbf{-xx}^{^{\prime
}}\right) +o\left( \varepsilon \right) \sim \frac{\varepsilon }{N}G\left( 
\mathbf{x}\right) ,
\end{eqnarray*}
since the deviation of $\mathbf{p}\left( \mathbf{x}\right) $ from $\mathbf{x}
$ induce second-order effects in $\varepsilon $ on the variance. This means 
\begin{equation}
\Delta \mathbf{x}_{t}=\mathbf{f}\left( \mathbf{x}_{t}\right) \varepsilon +%
\sqrt{\varepsilon }\frac{1}{\sqrt{N}}G^{1/2}\left( \mathbf{x}_{t}\right) 
\mathbf{\xi }_{t+\varepsilon },  \label{dse}
\end{equation}
where $\left( \mathbf{\xi }_{t};t\in \left\{ \varepsilon ,2\varepsilon
,...\right\} \right) $ is an iid $K-$dimensional Gaussian sequence with zero
mean and covariance matrix $I.$ For any $\delta >0$, using the multinomial
structure of the Wright-Fisher transition matrix, we have

\begin{equation*}
\underset{\varepsilon \rightarrow 0}{\lim }\frac{1}{\varepsilon }%
\int_{\left| \mathbf{y-x}\right| >\delta }\mathbf{P}\left( \mathbf{x}%
_{t+\varepsilon }\in d\mathbf{y\mid x}_{t}=\mathbf{x}\right) =0.
\end{equation*}
The stochastic dynamics (\ref{dse}) is a discretized version of the
diffusion process with continuous sample paths on $\mathcal{S}_{K}$ (making $%
\varepsilon \rightarrow 0$) 
\begin{equation}
d\mathbf{x}_{t}=\mathbf{f}\left( \mathbf{x}_{t}\right) dt+\frac{1}{\sqrt{N}}%
G^{1/2}\left( \mathbf{x}_{t}\right) d\mathbf{w}_{t},  \label{eq1}
\end{equation}
where $\mathbf{w}_{t}$ is a $K-$dimensional standard Brownian motion. Such
results on the diffusive approximation could be made rigorous using theorems
from \cite{EK}.

\subsubsection{\textbf{Speed (invariant) densities of the scaled process}}

The speed density of the diffusion process (\ref{eq1}), cancelling the
probability flux of its Kolmogorov forward (or Fokker-Planck) equation, is
(up to a multiplicative constant)

\begin{equation*}
m_{N}\left( \mathbf{x}\right) =\det \left( G^{-1}\left( \mathbf{x}\right)
\right) \exp 2N\int^{\mathbf{x}}\left( G^{-1}\mathbf{f}\right) \left( 
\mathbf{y}\right) \cdot d\mathbf{y,}
\end{equation*}
possibly not normalizable into a probability density function. If, for some
potential $V$, $\mathbf{f}\left( \mathbf{x}\right) =G(\mathbf{x})\nabla V(%
\mathbf{x}),$ then, by applying the gradient theorem to the line integral $%
\int^{\mathbf{x}}\left( G^{-1}\mathbf{f}\right) \left( \mathbf{y}\right)
\cdot d\mathbf{y}$ over some path in $\mathcal{S}_{K}$ ending in $\mathbf{x}$%
\begin{equation*}
m_{N}\left( \mathbf{x}\right) =\prod_{k=1}^{K}x_{k}^{-1}\exp 2NV(\mathbf{x}).
\end{equation*}
It may happen that, for some suitable choice of $V$, some normalizing
constant $Z_{N}$ turns $m_{N}$ to a probability density $p_{N}\left( \mathbf{%
x}\right) =Z_{N}^{-1}m_{N}\left( \mathbf{x}\right) $. In such cases, $%
p_{N}\left( \mathbf{x}\right) $ is the limiting equilibrium probability
distribution of some random variable $\mathbf{X}_{N}\sim p_{N}\left( \mathbf{%
x}\right) $ describing the asymptotic allelic frequencies. Note that a
non-integrable case is the neutral case with no drift $\mathbf{f}\left( 
\mathbf{x}\right) \equiv \mathbf{0}$ (or $\mathbf{p}\left( \mathbf{x}\right)
=\mathbf{x}$), with speed measure $m_{N}\left( \mathbf{x}\right)
=\prod_{k=1}^{K}x_{k}^{-1}$ on the simplex, but also $\mathbf{f}\left( 
\mathbf{x}\right) =G(\mathbf{x})\nabla V_{W}(\mathbf{x})$ as in (\ref{18c}),
with $V_{W}(\mathbf{x})=\frac{1}{2}\log \mathbf{x}^{\prime }W\mathbf{x,}$
and non-summable associated speed measure $m_{N}\left( \mathbf{x}\right)
=\left( \prod_{k=1}^{K}x_{k}^{-1}\right) \left( \mathbf{x}^{\prime }W\mathbf{%
x}\right) ^{N}$. We shall now consider situations including mutations where $%
m_{N}\left( \mathbf{x}\right) $ is integrable and thus normalizable.

\subsubsection{\textbf{The mutation/selection potential under the house of
cards condition for mutations}\protect\footnote{%
In the sequel, we consider a mean fitness $\mathbf{x}^{\prime }W\mathbf{x}$
but a frequency-dependent mean fitness $\mathbf{x}^{\prime }W\left( \mathbf{x%
}\right) \mathbf{x}$ would do as well provided $\mathbf{x}^{\prime }W\left( 
\mathbf{x}\right) \mathbf{x}$ is bounded as $\mathbf{x}\in S_{K}$.}}

For instance, if now $V(\mathbf{x})$ is the Shahshahani mutation/selection
potential under the house of cards condition for mutations (\ref{smpot}) 
\begin{equation}
V(\mathbf{x})=\log \mathcal{W}(\mathbf{x})\text{, where }\mathcal{W}(\mathbf{%
x})=e^{-\left| \mathbf{\mu }\right| }\prod_{k=1}^{K}x_{k}^{\mu _{k}}\left( 
\mathbf{x}^{\prime }W\mathbf{x}\right) ^{\left( 1-\left| \mathbf{\mu }%
\right| \right) /2},  \label{eq1a}
\end{equation}
then, with 
\begin{equation*}
Z_{N}\left( 2N\left| \mathbf{\mu }\right| \right) =\int_{\mathcal{S}_{K}}d%
\mathbf{x}\prod_{k=1}^{K}x_{k}^{2N\mu _{k}-1}\left( \mathbf{x}^{\prime }W%
\mathbf{x}\right) ^{N\left( 1-\left| \mathbf{\mu }\right| \right) }<\infty ,
\end{equation*}
\begin{equation}
p_{N}\left( \mathbf{x}\right) =\frac{1}{Z_{N}\left( 2N\left| \mathbf{\mu }%
\right| \right) }\prod_{k=1}^{K}x_{k}^{2N\mu _{k}-1}\left( \mathbf{x}%
^{\prime }W\mathbf{x}\right) ^{N\left( 1-\left| \mathbf{\mu }\right| \right)
}.  \label{eq2}
\end{equation}
If $\left| \mathbf{\mu }\right| =1$, we recognize the skew Dirichlet
distribution $D_{K}$ on the simplex $\mathcal{S}_{K},$ with parameters $%
2N\mu _{k}$, $k=1,...,K:$%
\begin{equation}
p_{N}\left( \mathbf{x}\right) =\frac{1}{Z_{N}^{D}\left( 2N\mathbf{\mu }%
\right) }\prod_{k=1}^{K}x_{k}^{2N\mu _{k}-1},\text{ }\mathbf{x\in }\mathcal{S%
}_{K},  \label{eq3}
\end{equation}
with normalizing partition function $Z_{N}^{D}\left( 2N\mathbf{\mu }\right)
=\prod_{k=1}^{K}\Gamma \left( 2N\mu _{k}\right) /\Gamma \left( 2N\left| 
\mathbf{\mu }\right| \right) $. We shall call $D_{K}\left( 2N\mathbf{\mu }%
\right) $ this distribution so that here $\left| \mathbf{\mu }\right|
=1\Rightarrow \mathbf{X}_{N}\mathbf{\sim }D_{K}\left( 2N\mathbf{\mu }\right) 
$. Except in the homogeneous case when all $\mu _{k}=\mu $ for each $%
k=1,...,K,$ Dirichlet distributed random variables with distribution (\ref
{eq3}) on the simplex are not exchangeable and $p_{N}\left( \mathbf{x}%
\right) $ is not invariant under a permutation of its coordinates, \cite
{King}. The same holds a fortiori for (\ref{eq2}) when $\left| \mathbf{\mu }%
\right| \neq 1$.

With $V$ as in (\ref{eq1a}), the deterministic dynamics corresponding to
this particular potential is thus $\Delta \mathbf{x}=\varepsilon G(\mathbf{x}%
)\nabla V(\mathbf{x})=:\varepsilon \mathbf{f}(\mathbf{x})$, as a version of (%
\ref{40pot}) when the time elapsed between two consecutive generations no
longer is $1$ but $\varepsilon $.

\subsubsection{\textbf{Weak mutation probabilities}}

If $2N\mu _{k}=\theta _{k}$ (small mutation probabilities, with all $\theta
_{k}>0$ now defining mutation rates), with $\mathbf{\theta }^{\prime
}:=\left( \theta _{1},...,\theta _{K}\right) $ and $\left| \mathbf{\theta }%
\right| :=\sum_{k=1}^{K}\theta _{k}$%
\begin{equation}
\mathbf{X}_{N}\mathbf{\sim }p_{N}\left( \mathbf{x}\right) =\frac{1}{%
Z_{N}\left( \mathbf{\theta }\right) }\prod_{k=1}^{K}x_{k}^{\theta
_{k}-1}\left( \mathbf{x}^{\prime }W\mathbf{x}\right) ^{N-\left| \mathbf{%
\theta }\right| /2},  \label{eq2w}
\end{equation}
where 
\begin{equation}
Z_{N}\left( \mathbf{\theta }\right) =\int_{\mathcal{S}_{K}}d\mathbf{x}%
\prod_{k=1}^{K}x_{k}^{\theta _{k}-1}\left( \mathbf{x}^{\prime }W\mathbf{x}%
\right) ^{N-\left| \mathbf{\theta }\right| /2}<\infty .  \label{eq4}
\end{equation}
In the asymmetric mutation case when $\mathbf{\theta }^{\prime }\neq \theta 
\mathbf{1}^{\prime }$ for some common $\theta >0$, the probability
distribution of $\mathbf{X}_{N}$ on the simplex is not exchangeable either.
For similar shapes of the equilibrium distribution in this weak mutation
setup, see \cite{Wri}, \cite{Li1}, \cite{Li2} and \cite{Zen}.

With $\mathcal{Z}_{N}\left( \mathbf{\theta }\right) :=Z_{N}\left( \mathbf{%
\theta }\right) /Z^{D}\left( \mathbf{\theta }\right) $, $Z^{D}\left( \mathbf{%
\theta }\right) :=\prod_{k=1}^{K}\Gamma \left( \theta _{k}\right) /\Gamma
\left( \left| \mathbf{\theta }\right| \right) $, (\ref{eq2w}) is also 
\begin{equation*}
p_{N}\left( \mathbf{x}\right) =\frac{1}{\mathcal{Z}_{N}\left( \mathbf{\theta 
}\right) Z^{D}\left( \mathbf{\theta }\right) }\prod_{k=1}^{K}x_{k}^{\theta
_{k}-1}\left( \mathbf{x}^{\prime }W\mathbf{x}\right) ^{N-\left| \mathbf{%
\theta }\right| /2},
\end{equation*}
where 
\begin{equation}
\mathcal{Z}_{N}\left( \mathbf{\theta }\right) =\frac{1}{Z^{D}\left( \mathbf{%
\theta }\right) }\int_{\mathcal{S}_{K}}\prod_{k=1}^{K}x_{k}^{\theta
_{k}-1}\left( \mathbf{x}^{\prime }W\mathbf{x}\right) ^{N-\left| \mathbf{%
\theta }\right| /2}=\mathbf{E}\left( \mathbf{S}^{\prime }W\mathbf{S}\right)
^{N-\left| \mathbf{\theta }\right| /2}.  \label{eq5}
\end{equation}
and $\mathbf{S\sim }D_{K}\left( \mathbf{\theta }\right) $ is Dirichlet
distributed.

\subsubsection{\textbf{Weak mutation and weak selection probabilities}}

If in addition $W=J-\frac{1}{N}\overline{W}$ for some new symmetric fitness
differential matrix $\overline{W}\geq \mathbf{0}$ (involving an order $%
N^{-1} $ correction to the neutral model $J$ of selection), then $\mathbf{x}%
^{\prime }W\mathbf{x=}1-N^{-1}\mathbf{x}^{\prime }\overline{W}\mathbf{x}$
and 
\begin{equation*}
\left( \mathbf{x}^{\prime }W\mathbf{x}\right) ^{N-\left| \mathbf{\theta }%
\right| /2}\sim e^{-\mathbf{x}^{\prime }\overline{W}\mathbf{x}}\text{ for
large }N,
\end{equation*}
so that 
\begin{equation*}
p\left( \mathbf{x}\right) =\frac{1}{Z\left( \mathbf{\theta }\right) }%
\prod_{k=1}^{K}x_{k}^{\theta _{k}-1}e^{-\mathbf{x}^{\prime }\overline{W}%
\mathbf{x}},
\end{equation*}
for some normalizing constant 
\begin{equation}
Z\left( \mathbf{\theta }\right) =\int_{\mathcal{S}_{K}}d\mathbf{x}%
\prod_{k=1}^{K}x_{k}^{\theta _{k}-1}e^{-\mathbf{x}^{\prime }\overline{W}%
\mathbf{x}}.  \label{eq6}
\end{equation}
With $Z^{D}\left( \mathbf{\theta }\right) :=\prod_{k=1}^{K}\Gamma \left(
\theta _{k}\right) /\Gamma \left( \left| \mathbf{\theta }\right| \right)
=B_{k}\left( \mathbf{\theta }\right) $ (the multidimensional beta function)
and $\mathcal{Z}\left( \mathbf{\theta }\right) :=Z\left( \mathbf{\theta }%
\right) /Z^{D}\left( \mathbf{\theta }\right) $, we have 
\begin{equation*}
p\left( \mathbf{x}\right) =\frac{1}{\mathcal{Z}\left( \mathbf{\theta }%
\right) Z^{D}\left( \mathbf{\theta }\right) }\prod_{k=1}^{K}x_{k}^{\theta
_{k}-1}e^{-\mathbf{x}^{\prime }\overline{W}\mathbf{x}},
\end{equation*}
where 
\begin{equation}
\mathcal{Z}\left( \mathbf{\theta }\right) =\frac{1}{Z^{D}\left( \mathbf{%
\theta }\right) }\int_{\mathcal{S}_{K}}\prod_{k=1}^{K}x_{k}^{\theta
_{k}-1}e^{-\mathbf{x}^{\prime }\overline{W}\mathbf{x}}=\mathbf{E}\left( e^{-%
\mathbf{S}^{\prime }\overline{W}\mathbf{S}}\right)  \label{eq7}
\end{equation}
with $\mathbf{S\sim }D_{K}\left( \mathbf{\theta }\right) .$ The latter $%
p\left( \mathbf{x}\right) $ is the unique invariant probability density of
the diffusion process on the simplex, \cite{THJ}, \cite{Ev} 
\begin{equation}
d\mathbf{x}_{t}=\mathbf{f}\left( \mathbf{x}_{t}\right) dt+G^{1/2}\left( 
\mathbf{x}_{t}\right) d\mathbf{w}_{t},  \label{eq8}
\end{equation}
where the drift $\mathbf{f}\left( \mathbf{x}\right) $ is 
\begin{equation*}
\begin{array}{l}
\mathbf{f}\left( \mathbf{x}\right) =G(\mathbf{x})\nabla \overline{V}(\mathbf{%
x})\text{ with} \\ 
\overline{V}(\mathbf{x})=\log \overline{\mathcal{W}}(\mathbf{x})\text{, and }%
\overline{\mathcal{W}}\left( \mathbf{x}\right) =\left(
\prod_{k=1}^{K}x_{k}^{\theta _{k}}\right) e^{-\mathbf{x}^{\prime }\overline{W%
}\mathbf{x}}
\end{array}
.
\end{equation*}
Namely, 
\begin{equation}
\begin{array}{l}
p\left( \mathbf{x}\right) =\frac{1}{Z\left( \mathbf{\theta }\right) }\det
\left( G^{-1}\right) \exp \int^{\mathbf{x}}\left( G^{-1}\mathbf{f}\right)
\left( \mathbf{y}\right) \cdot d\mathbf{y} \\ 
=\frac{1}{Z\left( \mathbf{\theta }\right) }\left(
\prod_{k=1}^{K}x_{k}^{-1}\right) \overline{\mathcal{W}}(\mathbf{x})=\frac{1}{%
Z\left( \mathbf{\theta }\right) }\prod_{k=1}^{K}x_{k}^{\theta _{k}-1}e^{-%
\mathbf{x}^{\prime }\overline{W}\mathbf{x}}
\end{array}
.  \label{eq9}
\end{equation}
A unit time $t$ in the latter diffusion process represents $N$ generations
of the discrete-time model. See \cite{Wri}, \cite{Li2}, \cite{Gri} and \cite
{Bar}.

\subsection{Computing the partition functions $Z_{N}\left( \mathbf{\theta }%
\right) $ and $Z\left( \mathbf{\theta }\right) $}

\subsubsection{\textbf{A constructive formula for computing with Dirichlet}$%
\left( \mathbf{\theta }\right) $\textbf{\ distribution}}

The following `Gamma-calculus' result will be useful (see \cite{HM1}):

\begin{theorem}
\label{Theo2} Consider an asymmetric Dirichlet distributed random variable
on the $\left( K-1\right) -$simplex, viz: $\mathbf{S}\overset{d}{\sim }$ $%
D_{K}\left( \mathbf{\theta }\right) $ and let $\mathbf{S}(t)=t\mathbf{S}(1)=t%
\mathbf{S=}\left( tS_{1},...,tS_{K}\right) $, $t>0$, with $%
\sum_{k=1}^{K}S_{k}(t)=t\mathbf{.}$

$(i)$ Let $f$ be any Borel-measurable function for which 
\begin{equation*}
\int_{0}^{\infty }\mathbf{E}\left( \left| f(\mathbf{S}(t))\right| \right)
t^{\left| \mathbf{\theta }\right| -1}e^{-pt}dt<\infty \text{, }p>0\text{.}
\end{equation*}
Then, with $\mathbf{T}(p):=(T_{k}(p);k=1,...,K)$, $K$ independent random
variables defined by $T_{k}(p)=\frac{1}{p}T_{k}$, $p>0$, $k=1,...,K$ where $%
T_{k}(1):=T_{k}\overset{d}{\sim }$ gamma$\left( \theta _{k}\right) $, we
have 
\begin{equation}
\int_{0}^{\infty }\mathbf{E}\left( f(\mathbf{S}(t))\right) t^{\left| \mathbf{%
\theta }\right| -1}e^{-pt}dt=\frac{\Gamma \left( \left| \mathbf{\theta }%
\right| \right) }{p^{\left| \mathbf{\theta }\right| }}\mathbf{E}\left( f(%
\mathbf{T}(p))\right) .  \label{eq18}
\end{equation}
$(ii)$ If $f$ is homogeneous of degree $d$, i.e. if $f(t\mathbf{s})=t^{d}f(%
\mathbf{s})$, $t>0$, $\mathbf{s}:=\left( s_{1},...,s_{K}\right) \in \mathbf{R%
}^{n}$, and if $\mathbf{E}\left( \left| f(\mathbf{S})\right| \right) <\infty 
$ then, with $\mathbf{T}^{\prime }:=\left( T_{1},...,T_{K}\right) $, 
\begin{equation}
\mathbf{E}\left( f(\mathbf{S})\right) =\frac{\Gamma \left( \left| \mathbf{%
\theta }\right| \right) }{\Gamma \left( \left| \mathbf{\theta }\right|
+d\right) }\mathbf{E}\left( f(\mathbf{T})\right) .  \label{eq19}
\end{equation}
\end{theorem}

This shows that computing the expected value of some functional with respect
to the asymmetric Dirichlet distribution on the simplex can be achieved
while averaging over an identically distributed gamma distributed sample; a
much simpler task.

\subsubsection{\textbf{Estimating }$Z_{N}\left( \mathbf{\theta }\right) $%
\textbf{\ in (\ref{eq4})}}

\begin{theorem}
When $N\gg \left| \mathbf{\theta }\right| /2$, 
\begin{equation}
Z_{N}\left( \mathbf{\theta }\right) \sim \frac{N!}{\Gamma \left( 2N\right) }%
\sum_{\left| \left[ \mathbf{n}\right] \right| =N}\prod_{k,l=1}^{K}\frac{%
W_{k,l}^{n_{k,l}}}{n_{k,l}!}\prod_{k=1}^{K}\Gamma \left( \theta _{k}+\alpha
_{k}\left( \left[ \mathbf{n}\right] \right) \right) ,  \label{eq10}
\end{equation}
where the summation runs over all integer ordered partitions $\left[ \mathbf{%
n}\right] :=n_{k,l}$ of $N$ as $\left| \left[ \mathbf{n}\right] \right|
:=\sum_{k,l=1}^{K}n_{k,l}=N$ and $\alpha _{k}\left( \left[ \mathbf{n}\right]
\right) =\sum_{l=1}^{K}\left( n_{k,l}+n_{l,k}\right) $.
\end{theorem}

\emph{Proof:} Consider the statement $(ii)$ of Theorem \ref{Theo2} with $f(%
\mathbf{S})=\left( \mathbf{S}^{\prime }\overline{W}\mathbf{S}\right)
^{N-\left| \mathbf{\theta }\right| /2}$. This function $f$ is homogeneous of
degree $d=2N-\left| \mathbf{\theta }\right| $ . Thus 
\begin{equation*}
\mathcal{Z}_{N}\left( \mathbf{\theta }\right) =\mathbf{E}\left( \mathbf{S}%
^{\prime }W\mathbf{S}\right) ^{N-\left| \mathbf{\theta }\right| /2}=\frac{%
\Gamma \left( \left| \mathbf{\theta }\right| \right) }{\Gamma \left(
2N\right) }\mathbf{E}\left( \mathbf{T}^{\prime }W\mathbf{T}\right)
^{N-\left| \mathbf{\theta }\right| /2}
\end{equation*}
where the random vector $\mathbf{T}$ has independent components with $T_{k}%
\overset{d}{\sim }$ gamma$\left( \theta _{k}\right) .$ Thus, with $\mathbf{t}%
^{\prime }=\left( t_{1},...,t_{K}\right) $%
\begin{eqnarray*}
\mathcal{Z}_{N}\left( \mathbf{\theta }\right) &=&\frac{\Gamma \left( \left| 
\mathbf{\theta }\right| \right) }{\prod_{k=1}^{K}\Gamma \left( \theta
_{k}\right) }\frac{1}{\Gamma \left( 2N\right) }\int_{\mathbf{R}%
_{+}^{K}}\left( \mathbf{t}^{\prime }W\mathbf{t}\right) ^{N-\left| \mathbf{%
\theta }\right| /2}\prod_{k=1}^{K}t_{k}^{\theta _{k}-1}e^{-t_{k}}dt_{k}\text{
or} \\
Z_{N}\left( \mathbf{\theta }\right) &=&\frac{1}{\Gamma \left( 2N\right) }%
\int_{\mathbf{R}_{+}^{K}}\left( \mathbf{t}^{\prime }W\mathbf{t}\right)
^{N-\left| \mathbf{\theta }\right| /2}\prod_{k=1}^{K}t_{k}^{\theta
_{k}-1}e^{-t_{k}}dt_{k}.
\end{eqnarray*}
When $N$ is large enough, $\mathcal{Z}_{N}\left( \mathbf{\theta }\right) $
is close to $\frac{\Gamma \left( \left| \mathbf{\theta }\right| \right) }{%
\Gamma \left( 2N\right) }\mathbf{E}\left( \mathbf{T}^{\prime }W\mathbf{T}%
\right) ^{N}$, with 
\begin{equation*}
\mathbf{E}\left( \mathbf{T}^{\prime }W\mathbf{T}\right) ^{N}=N!\sum_{\left|
\left[ \mathbf{n}\right] \right| =N}\prod_{k,l=1}^{K}\frac{W_{k,l}^{n_{k,l}}%
}{n_{k,l}!}\prod_{k=1}^{K}\mathbf{E}\left( T_{k}^{\alpha _{k}\left( \left[ 
\mathbf{n}\right] \right) }\right) .
\end{equation*}
In the latter expression, the summation runs over all square $K\times K$
arrays $\left[ \mathbf{n}\right] _{k;l}\mathbf{:=}n_{k,l}$ whose integral
non-negative entries sum to $N$ (the set $\left\{ \left[ \mathbf{n}\right]
:\left| \left[ \mathbf{n}\right] \right| =N\right\} $)\footnote{%
The number of ordered partitions of the integer $N$ into $k$ non-negative
integral summands, $k=1,...,K^{2}$, is $\binom{N+k-1}{k-1}$. There are thus $%
\sum_{k=1}^{K^{2}}\binom{N+k-1}{k-1}=$ $\binom{N+K^{2}-1}{K^{2}-1}\underset{N%
\text{ large}}{\sim }N^{K^{2}-1}/\left( K^{2}-1\right) !$ ways to realize $%
\left| \left[ \mathbf{n}\right] \right| :=\sum_{k,l=1}^{K}n_{k,l}=N$, with $%
n_{k,l}$ non-negative integers.}, and $\alpha _{k}\left( \left[ \mathbf{n}%
\right] \right) :=\sum_{l=1}^{K}\left( n_{k,l}+n_{l,k}\right) $, integers.
Recalling the moment structure of gamma random variables (rvs), $\mathbf{E}%
\left( T_{k}^{\alpha _{k}\left( \left[ \mathbf{n}\right] \right) }\right) =%
\frac{\Gamma \left( \theta _{k}+\alpha _{k}\left( \left[ \mathbf{n}\right]
\right) \right) }{\Gamma \left( \theta _{k}\right) }$, we thus get 
\begin{eqnarray*}
\mathcal{Z}_{N}\left( \mathbf{\theta }\right) &\sim &\frac{\Gamma \left(
\left| \mathbf{\theta }\right| \right) }{\Gamma \left( 2N\right) }%
N!\sum_{\left| \left[ \mathbf{n}\right] \right| =N}\prod_{k,l=1}^{K}\frac{%
W_{k,l}^{n_{k,l}}}{n_{k,l}!}\prod_{k=1}^{K}\frac{\Gamma \left( \theta
_{k}+\alpha _{k}\left( \left[ \mathbf{n}\right] \right) \right) }{\Gamma
\left( \theta _{k}\right) } \\
Z_{N}\left( \mathbf{\theta }\right) &\sim &\frac{N!}{\Gamma \left( 2N\right) 
}\sum_{\left| \left[ \mathbf{n}\right] \right| =N}\prod_{k,l=1}^{K}\frac{%
W_{k,l}^{n_{k,l}}}{n_{k,l}!}\prod_{k=1}^{K}\Gamma \left( \theta _{k}+\alpha
_{k}\left( \left[ \mathbf{n}\right] \right) \right) .\text{ }\Box
\end{eqnarray*}

\subsubsection{\textbf{Computing }$Z\left( \mathbf{\theta }\right) $\textbf{%
\ in (\ref{eq6})}}

We now show that

\begin{theorem}
\begin{equation}
\begin{array}{l}
Z\left( \mathbf{\theta }\right) =\frac{\prod_{k=1}^{K}\Gamma \left( \theta
_{k}\right) }{\Gamma \left( \left| \mathbf{\theta }\right| \right) }\times
\\ 
\left( 1+\frac{\Gamma \left( \left| \mathbf{\theta }\right| \right) }{%
\prod_{k=1}^{K}\Gamma \left( \theta _{k}\right) }\sum_{N\geq 1}\left(
-1\right) ^{N}\sum_{\left| \left[ \mathbf{n}\right] \right| =N}\frac{%
\prod_{k=1}^{K}\Gamma \left( \theta _{k}+\alpha _{k}\left( \left[ \mathbf{n}%
\right] \right) \right) }{\Gamma \left( \left| \mathbf{\theta }\right|
+2N\right) }\prod_{k,l=1}^{K}\frac{\overline{W}_{k,l}^{n_{k,l}}}{n_{k,l}!}%
\right) .
\end{array}
\label{eq11}
\end{equation}
\end{theorem}

\emph{Proof:} Consider the statement $(i)$ of Theorem \ref{Theo2}. From (\ref
{eq18}), the right hand-side quantity, namely 
\begin{equation*}
\Gamma \left( \left| \mathbf{\theta }\right| \right) p^{-\left| \mathbf{%
\theta }\right| }\mathbf{E}\left( f(\mathbf{T}(p))\right)
\end{equation*}
may be interpreted as the Laplace transform in the variable $p$ of $\mathbf{E%
}\left( f(\mathbf{S}(t))\right) t^{\left| \mathbf{\theta }\right| -1}$
appearing in the left-hand side. Inverting this Laplace transform and
putting $t=1$ yields $\mathbf{E}\left( f(\mathbf{S})\right) $. This can be
used to compute $\mathcal{Z}\left( \mathbf{\theta }\right) =\mathbf{E}\left(
e^{-\mathbf{S}^{\prime }\overline{W}\mathbf{S}}\right) $ with $f(\mathbf{S}%
)=e^{-\mathbf{S}^{\prime }\overline{W}\mathbf{S}}.$ We have 
\begin{equation*}
\mathbf{E}\left( e^{-\mathbf{T}^{\prime }\left( p\right) \overline{W}\mathbf{%
T}\left( p\right) }\right) =1+\sum_{N\geq 1}\frac{\left( -1\right) ^{N}}{N!}%
\mathbf{E}\left( \mathbf{T}^{\prime }\left( p\right) \overline{W}\mathbf{T}%
\left( p\right) ^{N}\right) ,
\end{equation*}
where, 
\begin{equation*}
\mathbf{E}\left( \mathbf{T}^{\prime }\left( p\right) \overline{W}\mathbf{T}%
\left( p\right) ^{N}\right) =N!\sum_{\left| \left[ \mathbf{n}\right] \right|
=N}\prod_{k,l=1}^{K}\frac{\overline{W}_{k,l}^{n_{k,l}}}{n_{k,l}!}%
\prod_{k=1}^{K}\mathbf{E}\left( T_{k}\left( p\right) ^{\alpha _{k}\left(
\left[ \mathbf{n}\right] \right) }\right) .
\end{equation*}
Recalling $\mathbf{E}\left( T_{k}\left( p\right) ^{\alpha _{k}\left( \left[ 
\mathbf{n}\right] \right) }\right) =\frac{\Gamma \left( \theta _{k}+\alpha
_{k}\left( \left[ \mathbf{n}\right] \right) \right) }{\Gamma \left( \theta
_{k}\right) }p^{-\alpha _{k}\left( \left[ \mathbf{n}\right] \right) }$, we
get 
\begin{equation*}
\mathbf{E}\left( e^{-\mathbf{T}^{\prime }\left( p\right) \overline{W}\mathbf{%
T}\left( p\right) }\right) =1+\sum_{N\geq 1}\left( -1\right)
^{N}\sum_{\left| \left[ \mathbf{n}\right] \right| =N}\prod_{k,l=1}^{K}\frac{%
\overline{W}_{k,l}^{n_{k,l}}}{n_{k,l}!}\prod_{k=1}^{K}\frac{\Gamma \left(
\theta _{k}+\alpha _{k}\left( \left[ \mathbf{n}\right] \right) \right) }{%
\Gamma \left( \theta _{k}\right) }p^{-\alpha _{k}\left( \left[ \mathbf{n}%
\right] \right) }.
\end{equation*}
Owing to $\sum_{k=1}^{K}\alpha _{k}\left( \left[ \mathbf{n}\right] \right)
=2N$, this is also 
\begin{equation*}
\mathbf{E}\left( e^{-\mathbf{T}^{\prime }\left( p\right) \overline{W}\mathbf{%
T}\left( p\right) }\right) =1+\sum_{N\geq 1}p^{-2N}\sum_{\left| \left[ 
\mathbf{n}\right] \right| =N}\left( -1\right) ^{N}\prod_{k,l=1}^{K}\frac{%
\overline{W}_{k,l}^{n_{k,l}}}{n_{k,l}!}\prod_{k=1}^{K}\frac{\Gamma \left(
\theta _{k}+\alpha _{k}\left( \left[ \mathbf{n}\right] \right) \right) }{%
\Gamma \left( \theta _{k}\right) }.
\end{equation*}
The right hand-side quantity of statement (\ref{eq18}) is thus 
\begin{equation*}
\Gamma \left( \left| \mathbf{\theta }\right| \right) p^{-\left| \mathbf{%
\theta }\right| }\mathbf{E}\left( f(\mathbf{T}(p))\right) =
\end{equation*}
\begin{equation*}
\Gamma \left( \left| \mathbf{\theta }\right| \right) p^{-\left| \mathbf{%
\theta }\right| }+\Gamma \left( \left| \mathbf{\theta }\right| \right)
\sum_{N\geq 1}p^{-\left( 2N+\left| \mathbf{\theta }\right| \right)
}\sum_{\left| \left[ \mathbf{n}\right] \right| =N}\left( -1\right)
^{N}\prod_{k,l=1}^{K}\frac{\overline{W}_{k,l}^{n_{k,l}}}{n_{k,l}!}%
\prod_{k=1}^{K}\frac{\Gamma \left( \theta _{k}+\alpha _{k}\left( \left[ 
\mathbf{n}\right] \right) \right) }{\Gamma \left( \theta _{k}\right) }.
\end{equation*}
It is the Laplace-Stieltjes transform of $\mathbf{E}\left( e^{-\mathbf{S}%
^{\prime }\left( t\right) \overline{W}\mathbf{S}\left( t\right) }\right)
t^{\left| \mathbf{\theta }\right| -1}$.

But $p^{-\left( 2N+\left| \mathbf{\theta }\right| \right) }$ is the
Laplace-Stieltjes transform of $t^{2N+\left| \mathbf{\theta }\right|
-1}/\Gamma \left( 2N+\left| \mathbf{\theta }\right| \right) $ and $%
p^{-\left| \mathbf{\theta }\right| }$ the one of $t^{\left| \mathbf{\theta }%
\right| -1}/\Gamma \left( \left| \mathbf{\theta }\right| \right) $.
Inverting the Laplace transform and evaluating the result at $t=1$ yields 
\begin{equation}
\begin{array}{l}
\mathcal{Z}\left( \mathbf{\theta }\right) :=\mathbf{E}\left( e^{-\mathbf{S}%
^{\prime }\overline{W}\mathbf{S}}\right) \\ 
=1+\sum_{N\geq 1}\left( -1\right) ^{N}\frac{\Gamma \left( \left| \mathbf{%
\theta }\right| \right) }{\Gamma \left( 2N+\left| \mathbf{\theta }\right|
\right) }\sum_{\left| \left[ \mathbf{n}\right] \right| =N}\prod_{k,l=1}^{K}%
\frac{\overline{W}_{k,l}^{n_{k,l}}}{n_{k,l}!}\prod_{k=1}^{K}\frac{\Gamma
\left( \theta _{k}+\alpha _{k}\left( \left[ \mathbf{n}\right] \right)
\right) }{\Gamma \left( \theta _{k}\right) } \\ 
=1+\frac{1}{Z^{D}\left( \mathbf{\theta }\right) }\sum_{N\geq 1}\left(
-1\right) ^{N}\sum_{\left| \left[ \mathbf{n}\right] \right| =N}\frac{%
\prod_{k=1}^{K}\Gamma \left( \theta _{k}+\alpha _{k}\left( \left[ \mathbf{n}%
\right] \right) \right) }{\Gamma \left( \left| \mathbf{\theta }\right|
+2N\right) }\prod_{k,l=1}^{K}\frac{\overline{W}_{k,l}^{n_{k,l}}}{n_{k,l}!}.
\end{array}
\label{eqZ}
\end{equation}
Recalling $Z\left( \mathbf{\theta }\right) =Z^{D}\left( \mathbf{\theta }%
\right) \mathcal{Z}\left( \mathbf{\theta }\right) $, we thus obtain 
\begin{equation*}
\begin{array}{l}
Z\left( \mathbf{\theta }\right) =Z^{D}\left( \mathbf{\theta }\right) \mathbf{%
E}\left( e^{-\mathbf{S}^{\prime }\overline{W}\mathbf{S}}\right) \\ 
=Z^{D}\left( \mathbf{\theta }\right) \left( 1+\frac{1}{Z^{D}\left( \mathbf{%
\theta }\right) }\sum_{N\geq 1}\left( -1\right) ^{N}\sum_{\left| \left[ 
\mathbf{n}\right] \right| =N}\frac{\prod_{k=1}^{K}\Gamma \left( \theta
_{k}+\alpha _{k}\left( \left[ \mathbf{n}\right] \right) \right) }{\Gamma
\left( \left| \mathbf{\theta }\right| +2N\right) }\prod_{k,l=1}^{K}\frac{%
\overline{W}_{k,l}^{n_{k,l}}}{n_{k,l}!}\right) .\text{ }\Box
\end{array}
\end{equation*}

\begin{corollary}
The alternating series expansion of $\mathcal{Z}\left( \mathbf{\theta }%
\right) $ in (\ref{eqZ}) is convergent.
\end{corollary}

\emph{Proof}: Let 
\begin{equation*}
u_{N}\left( \mathbf{\theta }\right) :=\frac{1}{Z^{D}\left( \mathbf{\theta }%
\right) }\sum_{\left| \left[ \mathbf{n}\right] \right| =N}\frac{%
\prod_{k=1}^{K}\Gamma \left( \theta _{k}+\alpha _{k}\left( \left[ \mathbf{n}%
\right] \right) \right) }{\Gamma \left( \left| \mathbf{\theta }\right|
+2N\right) }\prod_{k,l=1}^{K}\frac{\overline{W}_{k,l}^{n_{k,l}}}{n_{k,l}!}
\end{equation*}
so that $\mathcal{Z}\left( \mathbf{\theta }\right) =1+\sum_{N\geq 1}\left(
-1\right) ^{N}u_{N}\left( \mathbf{\theta }\right) $. Let $\overline{w}>0$ be
the largest of the $\overline{W}_{k,l}$s. Then $\prod_{k,l=1}^{K}\overline{W}%
_{k,l}^{n_{k,l}}<\overline{w}^{N}$ and 
\begin{equation*}
u_{N}\left( \mathbf{\theta }\right) <\overline{w}^{N}\frac{1}{Z^{D}\left( 
\mathbf{\theta }\right) }\sum_{\left| \mathbf{n}\right| =N}\frac{%
\prod_{k=1}^{K}\Gamma \left( \theta _{k}+2n_{k}\right) }{\Gamma \left(
\left| \mathbf{\theta }\right| +2N\right) }\prod_{k=1}^{K}\frac{1}{n_{k}!},
\end{equation*}
where $\mathbf{n=}\left( n_{1},...,n_{K}\right) ^{\prime }$ is now a vector
with nonnegative integral entries summing to $N$. With $\mathbf{S}\sim
D_{K}\left( \mathbf{\theta }\right) ,$ we now have 
\begin{equation*}
\frac{1}{Z^{D}\left( \mathbf{\theta }\right) }\frac{\prod_{k=1}^{K}\Gamma
\left( \theta _{k}+2n_{k}\right) }{\Gamma \left( \left| \mathbf{\theta }%
\right| +2N\right) }=\mathbf{E}\left( \prod_{k=1}^{K}S_{k}^{2n_{k}}\right)
<1,
\end{equation*}
which are the integral moments of order $2\mathbf{n}$ of a Dirichlet$\left( 
\mathbf{\theta }\right) $ random vector (see (\ref{m4a}) below). Thus, $%
u_{N}\left( \mathbf{\theta }\right) <\frac{\left( K\overline{w}\right) ^{N}}{%
N!}$ and $\mathcal{Z}\left( \mathbf{\theta }\right) $ is absolutely
convergent hence convergent. $\Box $

\emph{Remarks:} As $N$ gets large, the contribution of $u_{N}\left( \mathbf{%
\theta }\right) $ to $\mathcal{Z}\left( \mathbf{\theta }\right) $ becomes
rapidly smaller and smaller suggesting that only a few first terms of the
series-expansion of $\mathcal{Z}\left( \mathbf{\theta }\right) $ should lead
to a satisfactory approximation. When $N=1$, the first-order term $%
-u_{1}\left( \mathbf{\theta }\right) $ is 
\begin{equation*}
-u_{1}\left( \mathbf{\theta }\right) =\frac{-1}{Z^{D}\left( \mathbf{\theta }%
\right) }\frac{\prod_{k=1}^{K}\Gamma \left( \theta _{k}\right) }{\Gamma
\left( \left| \mathbf{\theta }\right| +2\right) }\sum_{k,l=1}^{K}\theta _{k}%
\overline{W}_{k,l}\theta _{l}=\frac{-1}{\left| \mathbf{\theta }\right|
\left( \left| \mathbf{\theta }\right| +1\right) }\mathbf{\theta }^{\prime }%
\overline{W}\mathbf{\theta .}\text{ }
\end{equation*}
Note finally that, with $\mathbf{\alpha }\left( \left[ \mathbf{n}\right]
\right) :=\left( \alpha _{1}\left( \left[ \mathbf{n}\right] \right)
,...,\alpha _{K}\left( \left[ \mathbf{n}\right] \right) \right) ^{\prime }$
and $\mathbf{\theta +\alpha }\left( \left[ \mathbf{n}\right] \right) =\theta
_{k}+\alpha _{k}\left( \left[ \mathbf{n}\right] \right) ;$ $k=1,...,K$, the
factors 
\begin{equation*}
Z^{D}\left( \mathbf{\theta +\alpha }\left( \left[ \mathbf{n}\right] \right)
\right) =\frac{\prod_{k=1}^{K}\Gamma \left( \theta _{k}+\alpha _{k}\left(
\left[ \mathbf{n}\right] \right) \right) }{\Gamma \left( \left| \mathbf{%
\theta }\right| +2N\right) },
\end{equation*}
appearing in the series expansion of $Z\left( \mathbf{\theta }\right) $ are
the normalizing constants of a $D_{K}\left( \mathbf{\theta +\alpha }\left(
\left[ \mathbf{n}\right] \right) \right) $ Dirichlet distribution $\Diamond $%
.

\section{Generalized Ewens sampling formulae}

\subsection{The weak mutation/selection potential case}

Let 
\begin{equation}
\mathbf{X}\sim p\left( \mathbf{x}\right) =\frac{1}{Z\left( \mathbf{\theta }%
\right) }\prod_{k=1}^{K}x_{k}^{\theta _{k}-1}e^{-\mathbf{x}^{\prime }%
\overline{W}\mathbf{x}}\text{,}  \label{eqx}
\end{equation}
whose support is the simplex $\mathcal{S}_{K}.$ We wish to consider sampling
problems within $\mathbf{X}$ (as a random partition of the unit interval $%
\left[ 0,1\right] $) describing the equilibrium distribution of the allelic
frequencies, \cite{Ra}. Such generalized Ewens sampling formulae (ESF) were
also considered in \cite{Wa1}-\cite{Wa3}, in this context.

\subsubsection{\textbf{Moments}}

We first consider the simpler problem of computing the moments of $\mathbf{X}
$. With $\mathbf{m}^{\prime }:=\left( m_{1},...,m_{K}\right) $ non-negative
integers, the $\mathbf{m-}$moments of this stationary distribution are: 
\begin{equation}
\mathbf{E}\left( \prod_{k=1}^{K}X_{k}^{m_{k}}\right) =\frac{\int_{\mathcal{S}%
_{K}}d\mathbf{x}\prod_{k=1}^{K}x_{k}^{\theta _{k}+m_{k}-1}e^{-\mathbf{x}%
^{\prime }\overline{W}\mathbf{x}}}{\int_{\mathcal{S}_{K}}d\mathbf{x}%
\prod_{k=1}^{K}x_{k}^{\theta _{k}-1}e^{-\mathbf{x}^{\prime }\overline{W}%
\mathbf{x}}}=\frac{\mathcal{Z}\left( \mathbf{\theta }+\mathbf{m}\right) }{%
\mathcal{Z}\left( \mathbf{\theta }\right) },  \label{eq12}
\end{equation}
requiring the previous computation of $\mathcal{Z}\left( \mathbf{\theta }%
\right) $. With $m=\sum_{k=1}^{K}m_{k}$, we find from Equation (\ref{eqZ}) 
\begin{equation}
\begin{array}{l}
\mathbf{E}\left( \prod_{k=1}^{K}X_{k}^{m_{k}}\right) = \\ 
\frac{1+\frac{\Gamma \left( \left| \mathbf{\theta }\right| +m\right) }{%
\prod_{k=1}^{K}\Gamma \left( \theta _{k}+m_{k}\right) }\sum_{N\geq 1}\left(
-1\right) ^{N}\sum_{\left| \left[ \mathbf{n}\right] \right| =N}\frac{%
\prod_{k=1}^{K}\Gamma \left( \theta _{k}+m_{k}+\alpha _{k}\left( \left[ 
\mathbf{n}\right] \right) \right) }{\Gamma \left( \left| \mathbf{\theta }%
\right| +m+2N\right) }\prod_{k,l=1}^{K}\frac{\overline{W}_{k,l}^{n_{k,l}}}{%
n_{k,l}!}}{1+\frac{\Gamma \left( \left| \mathbf{\theta }\right| \right) }{%
\prod_{k=1}^{K}\Gamma \left( \theta _{k}\right) }\sum_{N\geq 1}\left(
-1\right) ^{N}\sum_{\left| \left[ \mathbf{n}\right] \right| =N}\frac{%
\prod_{k=1}^{K}\Gamma \left( \theta _{k}+\alpha _{k}\left( \left[ \mathbf{n}%
\right] \right) \right) }{\Gamma \left( \left| \mathbf{\theta }\right|
+2N\right) }\prod_{k,l=1}^{K}\frac{\overline{W}_{k,l}^{n_{k,l}}}{n_{k,l}!}}.
\end{array}
\label{mom}
\end{equation}
Note that, from the identity 
\begin{equation*}
\mathbf{E}\left[ \left( \sum_{k=1}^{K}u_{k}X_{k}\right) ^{m}\right]
=\sum_{\left| \mathbf{m}\right| =m}\binom{m}{m_{1}...m_{K}}\mathbf{E}\left(
\prod_{k=1}^{K}X_{k}^{m_{k}}\right) \prod_{k=1}^{K}u_{k}^{m_{k}},
\end{equation*}
\begin{equation*}
\binom{m}{m_{1}...m_{K}}\mathbf{E}\left( \prod_{k=1}^{K}X_{k}^{m_{k}}\right)
=\left[ \prod_{k=1}^{K}u_{k}^{m_{k}}\right] \mathbf{E}\left[ \left(
\sum_{k=1}^{K}u_{k}X_{k}\right) ^{m}\right] ,
\end{equation*}
where $\left( \sum_{k=1}^{K}u_{k}X_{k}\right) ^{m}$ is homogeneous of degree 
$m.$ In particular 
\begin{equation*}
1=\mathbf{E}\left[ \left( X_{1}+...+X_{K}\right) ^{m}\right] =\sum_{\left| 
\mathbf{m}\right| =m}\binom{m}{m_{1}...m_{K}}\mathbf{E}\left(
\prod_{k=1}^{K}X_{k}^{m_{k}}\right) .
\end{equation*}
The marginal moments $\mathbf{E}\left( X_{l}^{m_{l}}\right) $ are obtained
from $\mathbf{E}\left( \prod_{k=1}^{K}X_{k}^{m_{k}}\right) $ in (\ref{mom}),
while considering $m_{k}=0$ except for $k=l.$

\subsubsection{\textbf{Generalized Ewens Sampling Formula}}

Take a random (uniform) sequential $m-$sample without replacement from $%
\mathbf{X\sim }p\left( \mathbf{x}\right) ,$ describing the random
equilibrium distribution of the allelic frequencies. Suppose there are $%
P_{K,m}=p$ distinct visited types of alleles in the process; suppose also
that $K_{1}=k_{1},...,K_{p}=k_{p}\in \left\{ 1,...,K\right\} ^{p}$ are the
types of the visited alleles and that $B_{K,m}\left( k_{1}\right)
=m_{1},..,B_{K,m}\left( k_{p}\right) =m_{p}$ are the number of visits to (or
hits of) alleles number $k_{1},...,k_{p}$, entailing $m_{1},..,m_{p}\geq 1$
and $m=m_{1}+...+m_{p}$. We let $\mathbf{\theta }+\mathbf{m}_{p}:=\theta
_{k_{q}}+m_{q}$; $q=1,...,p$ shifting only the $\theta _{k_{q}}-$entries of $%
\mathbf{\theta }$ by $m_{q}$ and leaving the other ones unchanged. From (\ref
{eq12}) and (\ref{eqZ}), we obtain:

\begin{theorem}
With $m_{1},..,m_{p}\geq 1$ summing to $m$ and $p\leq m\wedge K$, the
probability of such an occupancy event is 
\begin{equation}
\mathbf{P}\left( K_{1}=k_{1},...,K_{p}=k_{p};B_{K,m}\left( k_{1}\right)
=m_{1},..,B_{K,m}\left( k_{p}\right) =m_{p};P_{K,m}=p\right)  \label{eq13}
\end{equation}
\begin{equation*}
\begin{array}{l}
=\mathbf{E}\left( \prod_{q=1}^{p}X_{k_{q}}^{m_{q}}\right) =\frac{\mathcal{Z}%
\left( \mathbf{\theta }+\mathbf{m}_{p_{{}}}\right) }{\mathcal{Z}\left( 
\mathbf{\theta }\right) }:=\frac{Z^{D}\left( \mathbf{\theta }\right)
_{{}}^{-1}\int_{\mathcal{S}_{K}}d\mathbf{x}\prod_{q=1}^{p}x_{k_{q}}^{\theta
_{k_{q}}+m_{q}-1}\prod_{k\neq \left\{ k_{1},...,k_{p}\right\} }x_{k}^{\theta
_{k}-1}e^{-\mathbf{x}^{\prime }\overline{W}\mathbf{x}}}{Z^{D}\left( \mathbf{%
\theta }\right) _{{}}^{-1}\int_{\mathcal{S}_{K}}d\mathbf{x}%
\prod_{k=1}^{K}x_{k}^{\theta _{k}-1}e^{-\mathbf{x}^{\prime }\overline{W}%
\mathbf{x}}} \\ 
=\frac{1}{\mathcal{Z}\left( \mathbf{\theta }\right) }(1+\frac{\Gamma \left(
\left| \mathbf{\theta }\right| +m\right) }{\prod_{q=1}^{p}\Gamma \left(
\theta _{k_{q}}+m_{q}\right) \prod_{k\neq \left\{ k_{1},...,k_{p}\right\}
}\Gamma \left( \theta _{k}\right) }\times \\ 
\sum_{N\geq 1}\left( -1\right) ^{N}\sum_{\left| \left[ \mathbf{n}\right]
\right| =N}\frac{\prod_{q=1}^{p}\Gamma \left( \theta _{k_{q}}+m_{q}+\alpha
_{k_{q}}\left( \left[ \mathbf{n}\right] \right) \right) \prod_{k\neq \left\{
k_{1},...,k_{p}\right\} }\Gamma \left( \theta _{k}+\alpha _{k}\left( \left[ 
\mathbf{n}\right] \right) \right) }{\Gamma \left( \left| \mathbf{\theta }%
\right| +m+2N\right) }\prod_{k,l=1}^{K}\frac{\overline{W}_{k,l}^{n_{k,l}}}{%
n_{k,l}!})
\end{array}
\end{equation*}
where $\mathcal{Z}\left( \mathbf{\theta }\right) $ is given by (\ref{eqZ}).
\end{theorem}

Considering the marginal event ``$K_{1}=k_{1},...,K_{p}=k_{p};P_{K,m}=p$'',
it holds that 
\begin{eqnarray*}
&&\mathbf{P}\left( K_{1}=k_{1},...,K_{p}=k_{p};P_{K,m}=p\right) \\
&=&\sum_{\left| \mathbf{m}\right| =m}^{\prime }\mathbf{P}\left(
K_{1}=k_{1},...,K_{p}=k_{p};B_{K,m}\left( k_{1}\right)
=m_{1},..,B_{K,m}\left( k_{p}\right) =m_{p};P_{K,m}=p\right) ,
\end{eqnarray*}
where the $^{\prime }-$sum runs over all positive integers $\mathbf{m}%
=\left( m_{1},..,m_{p}\right) $ summing to $m$. There are $\binom{m-1}{p-1}$
terms in this sum. And summing the latter probability over $%
k_{1},...,k_{p}\in \left\{ 1,...,K\right\} ^{p}$ gives $\mathbf{P}\left(
P_{K,m}=p\right) .$

\subsubsection{\textbf{Generalized ESF in the `Kingman }$*-$\textbf{limit', }%
\protect\cite{King}}

Let $\theta _{k}\rightarrow 0$\ for each $k=1,...,K$\ and $K\rightarrow
\infty $\ while $\sum_{k=1}^{K}\theta _{k}=\left| \mathbf{\theta }\right| 
\overset{*}{\rightarrow }\gamma >0.$\ We call it the $*-$limit and we wish
to consider the generalized Ewens sampling formula in this limiting
situation corresponding to a case with infinitely many alleles per locus, 
\cite{Wa0}, \cite{Feng}, \cite{Zhou}.

For instance, take $\theta _{k}=\theta /k$, $\sum_{k=1}^{K}\theta
_{k}=\theta H_{K}$\ where $H_{K}=\sum_{k=1}^{K}k^{-1}$, with $\theta
\rightarrow 0$, $K\rightarrow \infty $\ and $\theta \log K\rightarrow \gamma
>0.$

With (\ref{zs1}) and (\ref{zs2}) mentioned below, we have

\begin{theorem}
In the $*-$limit, the probability of the occupancy event: ``$%
K_{1}=k_{1},...,K_{m}=k_{p};B_{m}\left( k_{1}\right) =m_{1},..,B_{m}\left(
k_{p}\right) =m_{p};P_{m}=p$'', with $m_{1},..,m_{p}\geq 1$\ summing to $m,$%
\ exists and is given by 
\begin{equation}
\begin{array}{l}
\mathbf{P}\left( K_{1}=k_{1},...,K_{m}=k_{p};B_{m}\left( k_{1}\right)
=m_{1},..,B_{m}\left( k_{p}\right) =m_{p};P_{m}=p\right) \\ 
=\frac{\mathcal{Z}^{*}\left( \mathbf{\theta }+\mathbf{m}_{p_{{}}}\right) }{%
\mathcal{Z}^{*}\left( \mathbf{\theta }\right) }
\end{array}
,  \label{eq14}
\end{equation}
where $\mathcal{Z}^{*}\left( \mathbf{\theta }+\mathbf{m}_{p_{{}}}\right) $
and $\mathcal{Z}^{*}\left( \mathbf{\theta }\right) $ are given by (\ref{zs2}%
) and (\ref{zs1}) respectively.
\end{theorem}

\emph{Proof: }In the $*-$limit, $1/Z^{D}\left( \mathbf{\theta }\right)
=\Gamma \left( \left| \mathbf{\theta }\right| \right) /\prod_{k=1}^{K}\Gamma
\left( \theta _{k}\right) \overset{*}{\sim }\Gamma \left( \left| \mathbf{%
\theta }\right| \right) \prod_{k=1}^{K}\theta _{k}\rightarrow 0.$

Consider first \emph{\ }$\mathcal{Z}\left( \mathbf{\theta }\right) $ and
split the sum \emph{\ }$\sum_{\left| \left[ \mathbf{n}\right] \right| =N}$
into $\sum_{\left| \left[ \mathbf{n}\right] \right| =N}^{1}:=\sum_{\left|
\left[ \mathbf{n}\right] \right| =N;\alpha _{k}\left( \left[ \mathbf{n}%
\right] \right) =0}$ and $\sum_{\left| \left[ \mathbf{n}\right] \right|
=N}^{2}:=\sum_{\left| \left[ \mathbf{n}\right] \right| =N;\alpha _{k}\left(
\left[ \mathbf{n}\right] \right) \neq 0},$ where $\left\{ \mathbf{n:}\left|
\left[ \mathbf{n}\right] \right| =N;\alpha _{k}\left( \left[ \mathbf{n}%
\right] \right) =0\right\} $ corresponds to those arrays $\left[ \mathbf{n}%
\right] $ summing to $N$ and with $k-$th row and $k-$th column equal to $0$.
We have

\begin{center}
$
\begin{array}{l}
\mathcal{Z}\left( \mathbf{\theta }\right) =1+\frac{\Gamma \left( \left| 
\mathbf{\theta }\right| \right) }{\prod_{k=1}^{K}\Gamma \left( \theta
_{k}\right) }\sum_{N\geq 1}\left( -1\right) ^{N}\sum_{\left| \left[ \mathbf{n%
}\right] \right| =N}\frac{\prod_{k=1}^{K}\Gamma \left( \theta _{k}+\alpha
_{k}\left( \left[ \mathbf{n}\right] \right) \right) }{\Gamma \left( \left| 
\mathbf{\theta }\right| +2N\right) }\prod_{k,l=1}^{K}\frac{\overline{W}%
_{k,l}^{n_{k,l}}}{n_{k,l}!} \\ 
=1+\frac{\Gamma \left( \left| \mathbf{\theta }\right| \right) }{%
\prod_{k=1}^{K}\Gamma \left( \theta _{k}\right) }\sum_{N\geq 1}\left(
-1\right) ^{N}(\sum_{\left| \left[ \mathbf{n}\right] \right| =N}^{1}\frac{%
\prod_{k=1}^{K}\Gamma \left( \theta _{k}\right) }{\Gamma \left( \left| 
\mathbf{\theta }\right| +2N\right) }\prod_{k,l=1}^{K}\frac{\overline{W}%
_{k,l}^{n_{k,l}}}{n_{k,l}!}+ \\ 
+\sum_{\left| \left[ \mathbf{n}\right] \right| =N}^{2}\frac{%
\prod_{k=1}^{K}\Gamma \left( \theta _{k}+\alpha _{k}\left( \left[ \mathbf{n}%
\right] \right) \right) }{\Gamma \left( \left| \mathbf{\theta }\right|
+2N\right) }\prod_{k,l=1}^{K}\frac{\overline{W}_{k,l}^{n_{k,l}}}{n_{k,l}!})
\end{array}
$%
\begin{equation}
\overset{\ast }{\rightarrow }\mathcal{Z}^{*}\left( \mathbf{\theta }\right)
:=1+\frac{\Gamma \left( \gamma \right) }{\Gamma \left( \gamma +2N\right) }%
\sum_{N\geq 1}\left( -1\right) ^{N}\sum_{\left| \left[ \mathbf{n}\right]
\right| =N;\alpha _{k}\left( \left[ \mathbf{n}\right] \right)
=0}\prod_{k,l=1}^{\infty }\frac{\overline{W}_{k,l}^{n_{k,l}}}{n_{k,l}!}.
\label{zs1}
\end{equation}
\end{center}

Consider now \emph{\ }$\mathcal{Z}\left( \mathbf{\theta }+\mathbf{m}%
_{p}\right) $ and split the sum \emph{\ }$\sum_{\left| \left[ \mathbf{n}%
\right] \right| =N}$ into the two parts $\sum_{\left| \left[ \mathbf{n}%
\right] \right| =N}^{1}:=\sum_{\left| \left[ \mathbf{n}\right] \right|
=N;\alpha _{k}\left( \left[ \mathbf{n}\right] \right) =0,k\neq \left\{
k_{1},...,k_{p}\right\} }$ and $\sum_{\left| \left[ \mathbf{n}\right]
\right| =N}^{2}:=\sum_{\left| \left[ \mathbf{n}\right] \right| =N;\alpha
_{k}\left( \left[ \mathbf{n}\right] \right) \neq 0,k\neq \left\{
k_{1},...,k_{p}\right\} },$ where $\left\{ \mathbf{n:}\left| \left[ \mathbf{n%
}\right] \right| =N;\alpha _{k}\left( \left[ \mathbf{n}\right] \right)
=0,k\neq \left\{ k_{1},...,k_{p}\right\} \right\} $ corresponds to those
arrays $\left[ \mathbf{n}\right] $ whose entries sum to $N$ and with $k-$th
row and $k-$th column equal to $0$, but only for those $k\neq \left\{
k_{1},...,k_{p}\right\} $. We have

\begin{center}
$
\begin{array}{l}
\mathcal{Z}\left( \mathbf{\theta }+\mathbf{m}_{p_{{}}}\right) =1+\frac{%
\Gamma \left( \left| \mathbf{\theta }\right| +m\right) }{\prod_{q=1}^{p}%
\Gamma \left( \theta _{k_{q}}+m_{q}\right) \prod_{k\neq \left\{
k_{1},...,k_{p}\right\} }\Gamma \left( \theta _{k}\right) }\times \\ 
\sum_{N\geq 1}\left( -1\right) ^{N}(\sum_{\left| \left[ \mathbf{n}\right]
\right| =N}^{1}\frac{\prod_{q=1}^{p}\Gamma \left( \theta
_{k_{q}}+m_{q}+\alpha _{k_{q}}\left( \left[ \mathbf{n}\right] \right)
\right) \prod_{k\neq \left\{ k_{1},...,k_{p}\right\} }\Gamma \left( \theta
_{k}\right) }{\Gamma \left( \left| \mathbf{\theta }\right| +m+2N\right) }%
\prod_{k,l=1}^{K}\frac{\overline{W}_{k,l}^{n_{k,l}}}{n_{k,l}!} \\ 
+\sum_{\left| \left[ \mathbf{n}\right] \right| =N}^{2}\frac{%
\prod_{q=1}^{p}\Gamma \left( \theta _{k_{q}}+m_{q}+\alpha _{k_{q}}\left(
\left[ \mathbf{n}\right] \right) \right) \prod_{k\neq \left\{
k_{1},...,k_{p}\right\} }\Gamma \left( \theta _{k}+\alpha _{k}\left( \left[ 
\mathbf{n}\right] \right) \right) }{\Gamma \left( \left| \mathbf{\theta }%
\right| +m+2N\right) }\prod_{k,l=1}^{K}\frac{\overline{W}_{k,l}^{n_{k,l}}}{%
n_{k,l}!})
\end{array}
$%
\begin{equation}
\begin{array}{l}
\overset{\ast }{\rightarrow }\mathcal{Z}^{*}\left( \mathbf{\theta }+\mathbf{m%
}_{p_{{}}}\right) :=1+\frac{\Gamma \left( \gamma +m\right) }{\Gamma \left(
\gamma +m+2N\right) }\sum_{N\geq 1}\left( -1\right) ^{N} \\ 
\sum_{\left| \left[ \mathbf{n}\right] \right| =N;\alpha _{k}\left( \left[ 
\mathbf{n}\right] \right) =0,k\neq \left\{ k_{1},...,k_{p}\right\} }\frac{%
\prod_{q=1}^{p}\Gamma \left( m_{q}+\alpha _{k_{q}}\left( \left[ \mathbf{n}%
\right] \right) \right) }{\prod_{q=1}^{p}\Gamma \left( m_{q}\right) }%
\prod_{k,l=1}^{\infty }\frac{\overline{W}_{k,l}^{n_{k,l}}}{n_{k,l}!}.
\end{array}
\label{zs2}
\end{equation}
\end{center}

The $*-$limit expression of the occupancy event generalizes the Ewens
partition distribution in the infinitely-many-alleles population genetics
model with symmetric selection/mutation, as studied in \cite{Han} and \cite
{H1}.

\subsubsection{\textbf{Marginal distributions and frequency spectrum}}

Let 
\begin{equation*}
p_{k}\left( x_{k}\right) :=\int d\left( \mathbf{x\setminus }x_{k}\right)
p\left( \mathbf{x}\right)
\end{equation*}
be the $k-$th marginal of $\mathbf{X}\sim p\left( \mathbf{x}\right) $ as
given by (\ref{eqx}). Then 
\begin{equation}
\frac{1}{K}\sum_{k=1}^{K}p_{k}\left( x\right) dx=\mathbf{E}\left( \frac{1}{K}%
\sum_{k=1}^{K}\mathbf{1}\left( X_{k}\in dx\right) \right) ,  \label{fs}
\end{equation}
as the frequency spectrum, is the density of alleles at equilibrium in a
neighborhood of $x$, see \cite{Ew1}, \cite{Eli}. It is a uniform mixture of
the marginals $p_{k}\left( x\right) $ whose precise computation is possible
but involved. Note that, from (\ref{mom}), the marginal moments of $X_{k}$
are known.

\subsection{The house-of-cards mutations potential case\textbf{\ }}

We finally consider sampling formulae when the multi-allelic population is
at equilibrium and subject only to mutation driving forces.

If $V(\mathbf{x})$ is the mutation potential under the house of cards
condition 
\begin{equation*}
V(\mathbf{x})=\log \mathcal{W}_{M}(\mathbf{x})\text{, where }\mathcal{W}%
_{M}\left( \mathbf{x}\right) =e^{-\left| \mathbf{\mu }\right|
}\prod_{k=1}^{K}x_{k}^{\mu _{k}},
\end{equation*}
then, with 
\begin{equation*}
Z_{N}^{D}\left( 2N\left| \mathbf{\mu }\right| \right) =\int_{\mathcal{S}%
_{K}}d\mathbf{x}\prod_{k=1}^{K}x_{k}^{2N\mu _{k}-1}=\frac{%
\prod_{k=1}^{K}\Gamma \left( 2N\mu _{k}\right) }{\Gamma \left( 2N\left| 
\mathbf{\mu }\right| \right) }<\infty ,
\end{equation*}
\begin{equation*}
p_{N}\left( \mathbf{x}\right) =\frac{1}{Z_{N}^{D}\left( 2N\left| \mathbf{\mu 
}\right| \right) }\prod_{k=1}^{K}x_{k}^{2N\mu _{k}-1},\text{ }\mathbf{x}\in 
\mathcal{S}_{K}
\end{equation*}
the Dirichlet distribution $D_{K}$ on the simplex $\mathcal{S}_{K},$ with
parameters $2N\mu _{k}$, $k=1,...,K.$

When dealing with weak mutations, the allelic equilibrium distribution is 
\begin{equation}
p\left( \mathbf{x}\right) =\frac{1}{Z^{D}\left( \mathbf{\theta }\right) }%
\prod_{k=1}^{K}x_{k}^{\theta _{k}-1},\text{ }\mathbf{x}\in \mathcal{S}_{K},
\label{dir}
\end{equation}
the asymmetric Dirichlet distribution $D_{K}\left( \mathbf{\theta }\right) $
on the simplex $\mathcal{S}_{K},$ with parameters $\theta _{k}$, $k=1,...,K$
(unless $\mathbf{\theta }^{\prime }=\theta \mathbf{1}^{\prime }$ for some
common $\theta >0,$ in which case $D_{K}\left( \mathbf{\theta }\right) $
boils down to the symmetric Dirichlet distribution $D_{K}\left( \theta
\right) $). When dealing with the potential solely arising from mutations
(i.e. when avoiding selection), the allelic equilibrium probability
distribution is thus the one of $\mathbf{S}\overset{d}{=}\mathbf{X\sim }%
D_{K}\left( \mathbf{\theta }\right) $, corresponding to Dirichlet spacings.
A much simpler situation than the previous one involving both mutation and
selection but still not so obvious (\cite{Ew1}, section $5.10$). We shall
derive some sampling formulae in this context.

\subsubsection{\textbf{The frequency of a typical allele at equilibrium}}

Pick at random an allele from $\mathbf{S\sim }D_{K}\left( \mathbf{\theta }%
\right) $. It has frequency $S_{k}$ with probability $S_{k}$ and so the
frequency of a size-biased picked allele at equilibrium is 
\begin{equation*}
f\left( S_{1},...,S_{K}\right) =\sum_{k=1}^{K}S_{k}^{2},
\end{equation*}
an homogeneous functional of degree $2$. Applying $\left( ii\right) $ of
Theorem \ref{Theo2} 
\begin{equation}
\mathbf{E}\left( \sum_{k=1}^{K}S_{k}^{2}\right) =\frac{\Gamma \left( \left| 
\mathbf{\theta }\right| \right) }{\Gamma \left( \left| \mathbf{\theta }%
\right| +2\right) }\mathbf{E}\left( \sum_{k=1}^{K}T_{k}^{2}\right) =\frac{1}{%
\left| \mathbf{\theta }\right| \left( \left| \mathbf{\theta }\right|
+1\right) }\sum_{k=1}^{K}\theta _{k}\left( \theta _{k}+1\right) .  \label{sb}
\end{equation}
The variance of $\sum_{k=1}^{K}S_{k}^{2}$ requires the computation of the
expected value of $f\left( S_{1},...,S_{K}\right) =\left(
\sum_{k=1}^{K}S_{k}^{2}\right) ^{2}=\sum_{k=1}^{K}S_{k}^{4}+2\sum_{1\leq
k_{1}<k_{2}\leq K}^{{}}S_{k_{1}}^{2}S_{k_{2}}^{2}$, as an homogeneous
functional of degree $4$, which is 
\begin{eqnarray*}
&&\frac{\Gamma \left( \left| \mathbf{\theta }\right| \right) }{\Gamma \left(
\left| \mathbf{\theta }\right| +4\right) }\left[ \sum_{k=1}^{K}\mathbf{E}%
\left( T_{k}^{4}\right) +2\sum_{1\leq k_{1}<k_{2}\leq K}^{{}}\mathbf{E}%
\left( T_{k_{1}}^{2}\right) \mathbf{E}\left( T_{k_{2}}^{2}\right) \right] \\
&=&\frac{\Gamma \left( \left| \mathbf{\theta }\right| \right) }{\Gamma
\left( \left| \mathbf{\theta }\right| +4\right) }\left[ \sum_{k=1}^{K}\frac{%
\Gamma \left( \theta _{k}+4\right) }{\Gamma \left( \theta _{k}\right) }%
+2\sum_{1\leq k_{1}<k_{2}\leq K}^{{}}\frac{\Gamma \left( \theta
_{k_{1}}+2\right) }{\Gamma \left( \theta _{k_{1}}\right) }\frac{\Gamma
\left( \theta _{k_{2}}+2\right) }{\Gamma \left( \theta _{k_{2}}\right) }%
\right] .
\end{eqnarray*}
The full law of $\sum_{k=1}^{K}S_{k}^{2}$ could be obtained while
considering $f\left( S_{1},...,S_{K}\right) =\exp \left( \lambda
\sum_{k=1}^{K}S_{k}^{2}\right) .$ This functional is no longer homogeneous
and $\left( i\right) $ of Theorem \ref{Theo2} should then be applied, with
some combinatorics involved which we skip.

\subsubsection{\textbf{Smallest and largest allelic frequencies from }$%
\mathbf{S\sim }D_{K}\left( \mathbf{\theta }\right) $}

We shall now use $\left( i\right) $ of Theorem \ref{Theo2} to compute the
joint distribution of the largest and smallest allelic frequencies in a $%
D_{K}\left( \mathbf{\theta }\right) $ distributed population at equilibrium.
Suppose $1\geq b>a\geq 0$ and consider the spacings' functional 
\begin{equation}
f\left( S_{1},...,S_{K}\right) =\prod_{k=1}^{K}\mathbf{1}\left( a<S_{k}\leq
b\right) \text{.}  \label{minmax}
\end{equation}
Then $\mathbf{E}f\left( S_{1},...,S_{K}\right) =\mathbf{P}\left( S_{\left(
K\right) }>a,S_{\left( 1\right) }\leq b\right) $ is the required
probability, assuming $S_{\left( 1\right) }>...>S_{\left( K\right) }$ to be
the order statistics of $\left( S_{1},...,S_{K}\right) $. The case $a=0$ ($%
b=1$) gives the probability $\mathbf{P}\left( S_{\left( 1\right) }\leq
b\right) ,$ respectively $\mathbf{P}\left( S_{\left( K\right) }>a\right) $.

From statement $(i)$ of Theorem \ref{Theo2} indeed, the quantity 
\begin{equation*}
\Gamma \left( \left| \mathbf{\theta }\right| \right) p^{-\left| \mathbf{%
\theta }\right| }\prod_{k=1}^{K}\mathbf{P}\left( a<T_{k}\left( p\right) \leq
b\right) =\Gamma \left( \left| \mathbf{\theta }\right| \right)
\prod_{k=1}^{K}\left[ \frac{1}{\Gamma \left( \theta _{k}\right) }%
\int_{a}^{b}t^{\theta _{k}-1}e^{-pt}dt\right]
\end{equation*}
interprets as the Laplace transform of $\mathbf{P}\left( S_{\left( K\right)
}\left( t\right) >a,S_{\left( 1\right) }\left( t\right) \leq b\right)
t^{\left| \mathbf{\theta }\right| -1}.$ Inverting this Laplace transform and
putting $t=1$ yields $\mathbf{P}\left( S_{\left( K\right) }>a,S_{\left(
1\right) }\leq b\right) $. From this, we obtain directly 
\begin{equation}
\mathbf{P}\left( S_{\left( K\right) }>a,S_{\left( 1\right) }\leq b\right) =%
\frac{\Gamma \left( \left| \mathbf{\theta }\right| \right) }{%
\prod_{k=1}^{K}\Gamma \left( \theta _{k}\right) }*_{k=1}^{K}h_{\theta
_{k}}\left( 1\right) ,  \label{m1}
\end{equation}
where $*_{k=1}^{K}h_{\theta _{k}}\left( 1\right) $ is the $K$-fold
convolution of the functions 
\begin{equation*}
t\rightarrow h_{\theta _{k}}\left( t\right) =t^{\theta _{k}-1}\mathbf{1}%
\left( b\geq t>a\right) ,k=1,...,K,
\end{equation*}
evaluated at $t=1$. If $b=1$, (\ref{m1}) gives the tail probability
distribution of $S_{\left( K\right) }$ whereas $a=1$ gives the probability
distribution of $S_{\left( 1\right) }.$

We now briefly show in outline that these tools are also useful in the
computation of simple sampling formulae.

\subsubsection{\textbf{Sampling and the Dirichlet multinomial distribution}}

Let $\left( U_{1},...,U_{m}\right) $ be $m$ iid uniform throws on $\mathbf{S}%
\sim D_{K}\left( \mathbf{\theta }\right) $. Let $\mathbf{B}:=\left(
B_{1},...,B_{K}\right) $ be an integral-valued random vector which counts
the number of visits to the different types of alleles in a $m-$sample.
Hence, if $K_{l}$ is the allele type which the $l-$th trial meets, then $%
B_{k}:=\sum_{l=1}^{m}\mathbf{1}\left( K_{l}=k\right) $, $k=1,...,K.$

With $\sum_{k=1}^{K}m_{k}=m$ and $\mathbf{m}:=\left( m_{1},...,m_{K}\right) $%
, conditionally given $\mathbf{S}$, we have the multinomial distribution: 
\begin{equation*}
\mathbf{P}\left( \mathbf{B}=\mathbf{m}\mid \mathbf{S}\right) =\frac{m!}{%
\prod_{k=1}^{K}m_{k}!}\prod_{k=1}^{K}S_{k}^{m_{k}}.
\end{equation*}
Averaging over $\mathbf{S}$ and applying $\left( ii\right) $ of Theorem \ref
{Theo2} to compute $\mathbf{E}\left( \prod_{k=1}^{K}S_{k}^{m_{k}}\right) $,
we find 
\begin{equation}
\mathbf{P}\left( \mathbf{B}=\mathbf{m}\right) =\mathbf{EP}\left( \mathbf{B}=%
\mathbf{m}\mid \mathbf{S}\right) =\frac{m!}{\prod_{k=1}^{K}m_{k}!}\frac{%
\prod_{k=1}^{K}\left[ \theta _{k}\right] _{m_{k}}}{\left[ \left| \mathbf{%
\theta }\right| \right] _{m}},  \label{m2}
\end{equation}
where $\left[ \theta \right] _{m}:=\theta \left( \theta +1\right) ...\left(
\theta +m-1\right) ,$ $k\geq 1$, $\left( \theta \right) _{0}:=1.$ This
distribution is known as the Dirichlet multinomial distribution.

Applying Bayes formula, the posterior distribution of $\mathbf{S}$ given $%
\mathbf{B}=\mathbf{m}$ is determined by its density at a point $\mathbf{s}$
on the simplex $\mathcal{S}_{K}$ as 
\begin{equation*}
f_{\mathbf{S}}\left( \mathbf{s}\mid \mathbf{B}=\mathbf{m}\right) =\frac{%
\Gamma \left( \left| \mathbf{\theta }\right| +m\right) }{\prod_{k=1}^{K}%
\Gamma \left( \theta _{k}+m_{k}\right) }\prod_{k=1}^{K}s_{k}^{\left( \theta
_{k}+m_{k}\right) -1}\text{, }\mathbf{s\in }\mathcal{S}_{K}.
\end{equation*}
This shows, as is well-known, that $\mathbf{S}\mid \mathbf{B}=\mathbf{m}$ $%
\overset{d}{\sim }$ $D_{K}\left( \mathbf{\theta }+\mathbf{m}\right) $, where 
$\mathbf{\theta }+\mathbf{m}=\left( \theta _{1}+m_{1},...,\theta
_{K}+m_{K}\right) $ is obtained by shifting $\mathbf{\theta }$. In
particular 
\begin{equation*}
\mathbf{E}\left( S_{k}\mid \mathbf{B}=\mathbf{m}\right) =\frac{\theta
_{k}+m_{k}}{\left| \mathbf{\theta }\right| +m},\text{ }k=1,...,K.
\end{equation*}

\subsubsection{\textbf{P\`{o}lya urn sequence}}

This suggests the following recursive approach to the sampling formula where
successive samples are now drawn from the corresponding iterative posterior
distributions. More specifically, let $\left( K_{1},...K_{m}\right) \in
\left\{ 1,...,K\right\} ^{m}$ be the types of the successive alleles thus
drawn. Then, 
\begin{equation*}
\mathbf{P}\left( K_{1}=k_{1}\right) =\mathbf{E}\left( \mathbf{P}\left(
K_{1}=k_{1}\right) \mid \mathbf{S}\right) =\mathbf{E}\left( S_{k_{1}}\right)
=\frac{\theta _{k_{1}}}{\left| \mathbf{\theta }\right| },
\end{equation*}
\begin{equation*}
\mathbf{P}\left( K_{2}=k_{2}\mid K_{1}\right) =\frac{\theta _{k_{2}}+\mathbf{%
1}\left( K_{1}=k_{2}\right) }{\left| \mathbf{\theta }\right| +1},...,
\end{equation*}
\begin{equation*}
\mathbf{P}\left( K_{m}=k_{m}\mid K_{1},...,K_{m-1}\right) =\frac{\theta
_{k_{m}}+\sum_{q=1}^{m-1}\mathbf{1}\left( K_{q}=k_{m}\right) }{\left| 
\mathbf{\theta }\right| +m-1}.
\end{equation*}
The joint distribution of $\left( K_{1},...,K_{m}\right) $ reads 
\begin{equation}
\begin{array}{l}
\mathbf{P}\left( K_{1}=k_{1},...,K_{m}=k_{m}\right) \\ 
=\frac{\theta _{k_{1}}}{\left| \mathbf{\theta }\right| }\prod_{q=1}^{m-1}%
\frac{\theta _{k_{q+1}}+\sum_{r=1}^{q}\mathbf{1}\left( k_{r}=k_{q+1}\right) 
}{\left| \mathbf{\theta }\right| +q}=\frac{\prod_{q=1}^{m}\left( \theta
_{k_{q}}+\sum_{r=1}^{q-1}\mathbf{1}\left( k_{r}=k_{q}\right) \right) }{%
\left[ \left| \mathbf{\theta }\right| \right] _{m}}.
\end{array}
\label{m3}
\end{equation}
The sequence $K_{1},...,K_{m}$ is a P\`{o}lya urn sequence.

\subsubsection{\textbf{Asymptotics of the occupancy vector}}

The joint conditional generating function of the full occupancy vector $%
\mathbf{B}$ reads 
\begin{equation*}
\mathbf{E}\left( \prod_{k=1}^{K}u_{k}^{B_{k}}\mid \mathbf{S}\right) =\left(
\sum_{k=1}^{K}u_{k}S_{k}\right) ^{m},
\end{equation*}
which is homogeneous with degree $d=m$ allowing to compute $\mathbf{E}\left(
\prod_{k=1}^{K}u_{k}^{B_{k}}\right) .$ Further, with $\overline{T}%
_{k}:=T_{k}/\sum_{k=1}^{K}T_{k}$, $T_{k}\sim $ gamma$\left( \theta
_{k}\right) $, $k=1,...,K$, as above, using independence between $\left( 
\overline{T}_{k},\text{ }k=1,...,K\right) $ and $\sum_{k=1}^{K}T_{k}\sim $%
gamma$\left( \left| \mathbf{\theta }\right| \right) $ and recalling $\left( 
\overline{T}_{k},\text{ }k=1,...,K\right) \sim $ $D_{K}\left( \mathbf{\theta 
}\right) $

\begin{center}
$
\begin{array}{l}
\mathbf{E}\left( \prod_{k=1}^{K}u_{k}^{B_{k}/m}\right) =\frac{\Gamma \left(
\left| \mathbf{\theta }\right| \right) }{\Gamma \left( \left| \mathbf{\theta 
}\right| +m\right) }\mathbf{E}\left[ \left(
\sum_{k=1}^{K}u_{k}^{1/m}T_{k}\right) ^{m}\right] \\ 
\underset{m\uparrow \infty }{\sim }\frac{\Gamma \left( \left| \mathbf{\theta 
}\right| \right) }{\Gamma \left( \left| \mathbf{\theta }\right| +m\right) }%
\mathbf{E}\left[ \left( \sum_{k=1}^{K}T_{k}\right) ^{m}\left( 1+\frac{1}{m}%
\sum_{k=1}^{K}\overline{T}_{k}\log u_{k}\right) ^{m}\right] \\ 
\underset{m\uparrow \infty }{\sim }\frac{\Gamma \left( \left| \mathbf{\theta 
}\right| \right) }{\Gamma \left( \left| \mathbf{\theta }\right| +m\right) }%
\mathbf{E}\left( \sum_{k=1}^{K}T_{k}\right) ^{m}\mathbf{E}\left(
\prod_{k=1}^{K}u_{k}^{\overline{T}_{k}}\right) =\mathbf{E}\left(
\prod_{k=1}^{K}u_{k}^{\overline{T}_{k}}\right) =\mathbf{E}\left(
\prod_{k=1}^{K}u_{k}^{S_{k}}\right) ,
\end{array}
$
\end{center}

because $\mathbf{E}\left( \sum_{k=1}^{K}T_{k}\right) ^{m}=\frac{\Gamma
\left( \left| \mathbf{\theta }\right| +m\right) }{\Gamma \left( \left| 
\mathbf{\theta }\right| \right) }$. This shows that 
\begin{equation}
\mathbf{B}/m\overset{d}{\rightarrow }\mathbf{S}\text{ as }m\rightarrow
\infty .  \label{m4}
\end{equation}
Note that, applying the strong law of large numbers (conditionally given $%
\mathbf{S}$), the above convergence in law also holds almost surely: the
normalized occupancy vector $\mathbf{B}$ from an $m-$sample converges to $%
\mathbf{S}$ itself.\newline

\subsubsection{\textbf{ESF from equilibrium distribution driven solely by
mutations}}

Firstly, with $q_{k}>-\theta _{k}$, as we already observed, it holds 
\begin{equation}
\mathbf{E}\left( \prod_{k=1}^{K}S_{k}^{q_{k}}\right) =\frac{\Gamma \left(
\left| \mathbf{\theta }\right| \right) }{\Gamma \left( \left| \mathbf{\theta 
}\right| +\sum_{k=1}^{K}q_{k}\right) }\prod_{k=1}^{K}\frac{\Gamma \left(
\theta _{k}+q_{k}\right) }{\Gamma \left( \theta _{k}\right) }=\frac{%
Z^{D}\left( \mathbf{\theta }+\mathbf{q}\right) }{Z^{D}\left( \mathbf{\theta }%
\right) },  \label{m4a}
\end{equation}
where $\frac{\Gamma \left( \theta _{k}+q_{k}\right) }{\Gamma \left( \theta
_{k}\right) }=\left[ \theta _{k}\right] _{q_{k}}$ if $q_{k}$ is an integer.
Therefore, considering the probability of an occupancy event as in (\ref
{eq13}):

\begin{proposition}
With $m_{1},..,m_{p}\geq 1$ summing to $m$, the Ewens sampling formula under
asymmetric mutations only is
\end{proposition}

\begin{equation}
\begin{array}{l}
\mathbf{P}\left( K_{1}=k_{1},...,K_{p}=k_{p};B_{K,m}\left( k_{1}\right)
=m_{1},..,B_{K,m}\left( k_{p}\right) =m_{p};P_{K,m}=p\right) \\ 
=\mathbf{E}\left( \prod_{q=1}^{p}S_{k_{q}}^{m_{q}}\right) =\frac{Z^{D}\left( 
\mathbf{\theta }+\mathbf{m}_{p}\right) }{Z^{D}\left( \mathbf{\theta }\right) 
}=\frac{\Gamma \left( \left| \mathbf{\theta }\right| \right) }{\Gamma \left(
\left| \mathbf{\theta }\right| +m\right) }\prod_{q=1}^{p}\frac{\Gamma \left(
\theta _{k_{q}}+m_{q}\right) }{\Gamma \left( \theta _{k_{q}}\right) }
\end{array}
,  \label{m5}
\end{equation}
where $\frac{\Gamma \left( \theta _{k_{q}}+m_{q}\right) }{\Gamma \left(
\theta _{k_{q}}\right) }=\left[ \theta _{k_{q}}\right] _{m_{q}}.$

If we now consider the $*-$limit for mutations, 
\begin{equation}
\mathbf{E}\left( \prod_{q=1}^{p}S_{k_{q}}^{m_{q}}\right) \overset{*}{\sim }%
\frac{\Gamma \left( \gamma \right) }{\Gamma \left( \gamma +m\right) }%
\prod_{q=1}^{p}\theta _{k_{q}}\left( m_{q}-1\right) !\text{ }\overset{*}{%
\rightarrow }0,  \label{m6}
\end{equation}
showing that there is no proper $*-$limit of the occupancy probability (\ref
{m5}).\newline

\emph{Remark (standard Ewens sampling formula):} Suppose $\theta _{k}=\theta 
$, $k=1,...,K$ (the symmetric Dirichlet mutation model). Owing to
exchangeability of the alleles 
\begin{equation*}
\begin{array}{l}
\mathbf{P}\left( B_{K,m}\left( 1\right) =m_{1},..,B_{K,m}\left( p\right)
=m_{p};P_{K,m}=p\right) \\ 
=\binom{K}{p}\binom{m}{m_{1}...m_{p}}\frac{\Gamma \left( \left| \mathbf{%
\theta }\right| \right) }{\Gamma \left( \left| \mathbf{\theta }\right|
+m\right) }\prod_{q=1}^{p}\frac{\Gamma \left( \theta +m_{q}\right) }{\Gamma
\left( \theta \right) }
\end{array}
,
\end{equation*}
where the occupancy vector is, say, over the first $p$ alleles of $\mathcal{S%
}_{K}$.

Consider the $*-$limit where $\theta \rightarrow 0$, $K\rightarrow \infty $
while $\theta K\rightarrow \gamma >0$, \cite{King}. Owing to $\binom{K}{p}%
\overset{*}{\sim }K^{p}/p!,$ $\prod_{q=1}^{p}\theta _{k_{q}}=\theta ^{p}$ et 
$\left( K\theta \right) ^{p}\overset{*}{\sim }\gamma ^{p}$, we get 
\begin{equation}
\mathbf{P}\left( B_{K,m}\left( 1\right) =m_{1},..,B_{K,m}\left( p\right)
=m_{p};P_{K,m}=p\right) \overset{*}{\rightarrow }\frac{m!}{p!}\frac{\gamma
^{p}}{\left[ \gamma \right] _{m}}\frac{1}{\prod_{q=1}^{p}m_{q}}  \label{m7}
\end{equation}
which is formula $(29)$ of \cite{H0}, for example. In contrast with the
asymmetric Dirichlet model, the symmetric Dirichlet model admits a proper $%
*- $limit occupancy probability. This is one of the many facets\footnote{%
In \cite{H0}, the ESF (\ref{m7}) is called the first ESF. A second ESF
rather deals with the occupancy vector $\mathcal{A}_{K,m}\left( i\right) $, $%
i\in \left\{ 0,..,m\right\} $, which counts the number of alleles in the $m-$%
sample with $i$ representatives.} of the standard ESF, see \cite{H0} and 
\cite{Feng}. $\Diamond $\newline

Coming back to (\ref{m5}) in the asymmetric Dirichlet case, this suggests to
consider the following $**-$limit: choose $\left\{ \theta _{k}\right\} $ in
such a way that $\left| \mathbf{\theta }\right| :=\sum_{k=1}^{K}\theta
_{k}\rightarrow \gamma $ as $K\rightarrow \infty $, with none of the $\theta
_{k}\rightarrow 0$. Such a limiting model for the mutation rates $\theta
_{k} $ was considered in \cite{HM2}.

\emph{Examples are:}

$\left( i\right) $ $\theta _{k}=p^{k}$, $k=1,...,K$, for some $p\in \left(
0,1\right) $ with $\gamma =p/\left( 1-p\right) .$

$\left( ii\right) $ $\theta _{k}=k^{-2}$, $k=1,...,K$ with $\gamma =\zeta
\left( 2\right) =\pi ^{2}/6.$ $\Diamond $

In the $**-$limit, therefore 
\begin{equation}
\mathbf{E}\left( \prod_{q=1}^{p}S_{k_{q}}^{m_{q}}\right) \overset{**}{%
\rightarrow }\frac{\Gamma \left( \gamma \right) }{\Gamma \left( \gamma
+m\right) }\prod_{q=1}^{p}\frac{\Gamma \left( \theta _{k_{q}}+m_{q}\right) }{%
\Gamma \left( \theta _{k_{q}}\right) }.  \label{m8}
\end{equation}

The asymmetric Dirichlet model has a proper $**-$limit occupancy probability
(\ref{m5}), as given by (\ref{m8}).

To summarize:

\begin{theorem}
Consider the $m-$sampling problem from $\mathbf{S}\sim D_{K}\left( \mathbf{%
\theta }\right) $. With $m_{1},..,m_{p}\geq 1$ summing to $m$ and $p\leq
m\wedge K$, the probability of the occupancy event 
\begin{equation*}
``K_{1}=k_{1},...,K_{p}=k_{p};B_{K,m}\left( k_{1}\right)
=m_{1},..,B_{K,m}\left( k_{p}\right) =m_{p};P_{K,m}=p"
\end{equation*}

is given by (\ref{m5}). From (\ref{m7}) there is no non-degenerate limit of
this probability in the $*-$limit but there is one in the $**-$limit, given
by (\ref{m8}).
\end{theorem}

\subsubsection{\textbf{Moments, marginal distributions and frequency spectrum%
}}

Recalling (\ref{m4a}), the marginal moments of $S_{k}$ are $\mathbf{E}\left[
S_{k}^{q_{k}}\right] =\frac{\Gamma \left( \left| \mathbf{\theta }\right|
\right) }{\Gamma \left( \left| \mathbf{\theta }\right| +q_{k}\right) }\left[
\theta _{k}\right] _{q_{k}}$. So $S_{k}\sim $beta$\left( \theta _{k},\left| 
\mathbf{\theta }\right| -\theta _{k}\right) $ with marginal density $%
p_{k}\left( s_{k}\right) :=\int d\left( \mathbf{s\setminus }s_{k}\right)
p\left( \mathbf{s}\right) $ equal to 
\begin{equation*}
p_{k}\left( s\right) =\frac{\Gamma \left( \left| \mathbf{\theta }\right|
\right) }{\Gamma \left( \theta _{k}\right) \Gamma \left( \left| \mathbf{%
\theta }\right| -\theta _{k}\right) }s^{\theta _{k}-1}\left( 1-s\right)
^{\left| \mathbf{\theta }\right| -\theta _{k}-1},\text{ }s\in \left(
0,1\right) .
\end{equation*}
This gives the empirical average of $\mathbf{S}$ in a neighborhood of $s$ as 
\begin{equation*}
\mathbf{E}\left( \frac{1}{K}\sum_{k=1}^{K}\mathbf{1}\left( S_{k}\in
ds\right) \right) =\frac{1}{K}\sum_{k=1}^{K}p_{k}\left( s\right) ds.
\end{equation*}
Else, $f_{K}\left( s\right) =\frac{1}{K}\sum_{k=1}^{K}p_{k}\left( s\right) $%
, as the frequency spectrum, is the density of alleles at equilibrium in a
neighborhood of $s.$ It is a uniform mixture of beta$\left( \theta
_{k},\left| \mathbf{\theta }\right| -\theta _{k}\right) $ distributed rvs, $%
k=1,...,K$. Near $s=\left\{ 0,1\right\} $, $f_{K}\left( s\right) \sim
s^{\theta _{*}-1}$ and $f_{K}\left( s\right) \sim \left( 1-s\right) ^{\left| 
\mathbf{\theta }\right| -\theta _{*}-1}$, where $\theta _{*}=\min \left( 
\mathbf{\theta }\right) $ and $\theta ^{*}=\max \left( \mathbf{\theta }%
\right) .$

\textbf{Acknowledgments:} T. Huillet acknowledges partial support both from
the ``Chaire \textit{Mod\'{e}lisation math\'{e}matique et biodiversit\'{e}''}
and the labex MME-DII Center of Excellence (\textit{Mod\`{e}les
math\'{e}matiques et \'{e}conomiques de la dynamique, de l'incertitude et
des interactions}, ANR-11-LABX-0023-01 project).

\end{document}